\documentclass[letterpaper,twocolumn,10pt]{article}
\usepackage{jsys_camera_ready}

\microtypecontext{spacing=nonfrench}

\usepackage{tikz}
\usepackage{amsmath}
\usepackage{subfigure}
\usepackage[inline]{enumitem}
\usepackage[nolist]{acronym}
\usepackage[normalem]{ulem}
\usepackage{algorithm}
\usepackage{algpseudocode}
\usepackage{lipsum}
\usepackage{multirow}
\usepackage{pifont}
\usepackage{rotating}
\usepackage{xspace}
\usepackage[mode=text,detect-all,per-mode=symbol]{siunitx}
\usepackage{tikz}
\usepackage{color}

\definecolor{lightblue}{HTML}{2596be}

\definecolor{gray1}{gray}{.95}
\definecolor{gray2}{gray}{.85}
\definecolor{gray3}{gray}{.50}
\definecolor{darkgrey}{RGB}{54,54,54}
\definecolor{lightgrey}{RGB}{102,102,102}

\newcolumntype{H}{>{\setbox0=\hbox\bgroup}c<{\egroup}@{}}

\newcommand{\secref}[1]{Sec.~\ref{#1}}
\newcommand{\appdxref}[1]{Appx.~\ref{#1}}
\newcommand{\tabref}[1]{Tab.~\ref{#1}}
\newcommand{\figref}[1]{Fig.~\ref{#1}}

\newcommand{\cf}{cf.\xspace}
\newcommand{\ie}{i.e.,\xspace}
\newcommand{\eg}{e.g.,\xspace}
\newcommand{\Eg}{E.g.,\xspace}
\newcommand{\wrt}{w.r.t.\xspace}

\newcommand{\etal}{et~al.\xspace}

\newcommand*\circled[1]{\tikz[baseline=(char.base)]{
		\node[shape=circle,draw,inner sep=.5pt] (char) {#1};}}

\usepackage{wasysym}

\newcommand{\good}{$\CIRCLE$}

\usepackage{caption}
\begin{acronym}
	\acro{BN}{Bayesian Network}
	\acro{BLSTM}{Bidirectional Long Short Term Memory}
	\acro{DNN}{Deep Neural Network}
	\acro{DTMC}{Discrete-time Markov Chain}
	\acro{LSTM}{Long Short-Term Memory}
	\acro{GMM}{Gaussian Mixture Model}
	\acro{NN}{Neural Network}
	\acro{PST}{Probabilistic Suffix Tree}
	\acro{RF}{Random Forest}
	\acro{SVD}{Singular Value Decomposition}
	\acro{SVM}{Support Vector Machine}
	\acro{TA}{Timed Automata}
	\acro{PCA}{Principal Component Analysis}

	\acro{CPS}{cyber-physical system}
	\acro{HMI}{Human-Machine Interface}
	\acro{PLC}{Programmable Logic Controller}
	\acro{RTU}{Remote Terminal Unit}
	\acro{MTU}{Master Terminal Unit}
	\acro{ICS}{Industrial Control System}
	\acro{IDS}{Intrusion Detection System}
	\acro{IIDS}{Industrial Intrusion Detection System}
	\acro{SCADA}{Supervisory Control and Data Acquisition}

	\acro{DoS}{Denial of Service}
	\acro{DPI}{Deep Packet Inspection}
	\acro{IP}{Internet Protocol}
	\acro{MitM}{Machine-in-the-Middle}

	\acro{SWaT}{Secure Water Treatment}
	\acro{TEP}{Tennessee Eastman Process}
	\acro{WADI}{Water Distribution}
	\acro{CISS}{Critical Infrastructure Security Showdown}

	\acro{PASAD}{Process-Aware Stealthy Attack Detector}
	\acro{TABOR}{Timed Automata and Bayesian netwORk}
	\acro{OOA}{Out Of Alphabet}
	\acro{SWIDE}{Sliding WIndow based on Differential sEgmentation}

	\acro{TPR}{True Positive Rate}
	\acro{FPR}{False Positive Rate}
	\acro{ODR}{Overall Detection Rate}

	\acro{IPAL}{Industrial Protocol Abstraction Layer}
	\acro{SLR}{Systematic Literature Review}
	\acro{SMS}{Systematic Mapping Study}
	\acro{SoK}{Systematization of Knowledge}
\end{acronym}
\newcommand{ \evalAcceptedPaper }{ 609 }
\newcommand{ \evalMetricsUsedAtLeastTenTimes }{ 12 }

\newcommand{ \evalPapersWithRecall }{ 265 }

\newcommand{ \evalPapersWithPrecision }{ 162 }
\newcommand{ \evalPapersWithFScore }{ 153 }
\newcommand{ \evalPapersWithTimeaware }{ 16 }

\newcommand{ \evalPapersWithConfusionMatrix }{ 59 }

\newcommand{ \evalPapersWithDetectionDelay }{ 48 }

\newcommand{ \evalAverageCiteCount }{ 2.9 }
\newcommand{ \evalCitationGraphOmitted }{ 125 }
\newcommand{ \evalGlobalClusteringCoefficient }{ 0.11 }

\newcommand{ \evalProcessDatasetPapers }{ 102 }
\newcommand{ \evalProcessClusteringCoefficient }{ 0.15 }

\newcommand{ \evalNetworkDatasetPapers }{ 81 }
\newcommand{ \evalNetworkClusteringCoefficient }{ 0.13 }

\newcommand{ \evalBothDatasetPapers }{ 19 }

\newcommand{ \evalComparesToOnAvg }{ 0.5 }
\newcommand{ \evalCouldCompareToOnAvg }{ 6.0 }
\newcommand{ \evalSWaTComparesToOnAvg }{ 2.4 }
\newcommand{ \evalSWaTCouldCompareToOnAvg }{ 10.0 }
\newcommand{ \evalMorrisGasComparesToOnAvg }{ 1.7 }
\newcommand{ \evalMorrisGasCouldCompareToOnAvg }{ 16.1 }
\newcommand{ \evalNotComparedToASinglePaper }{ 95.4 }
\newcommand{ \evalAverageDatasetsPerPaper }{ 1.3 }
\newcommand{ \evalPaperWithOneDataset }{ 509 }

\newcommand{ \evalPaperWithMultipleDatasets }{ 100 }
\newcommand{ \evalPercentagePaperWithMultipleDatasets }{ 16.4 }
\newcommand{ \evalUniquePublicDatasets }{ 35 }

\newcommand{ \evalSWaTUsedInPaperPercent }{ 9.0 }

\newcommand{ \evalBATADALUsedInPaperPercent }{ 1.6 }

\newcommand{ \evalWADIUsedInPaperPercent }{ 1.0 }

\newcommand{ \evalMorrisGasUsedInPaperPercent }{ 11.8 }

\newcommand{ \evalMorrisPowerUsedInPaperPercent }{ 5.6 }

\newcommand{ \evalMorrisWaterUsedInPaperPercent }{ 2.8 }

\newcommand{ \evalWaterPlantUsedInPaperPercent }{ 2.0 }

\newcommand{ \evalHAIUsedInPaperPercent }{ 1.1 }

\newcommand{ \evalLemayUsedInPaperPercent }{ 1.0 }
\newcommand{ \evalSWaTMorrisGasPercentage }{ 20.4 }

\newcommand{ \evalPapersWithPublicDatasetPercent }{ 33.3 }

\newcommand{ \evalDatasetsUsedOnce }{ 16 }
\newcommand{ \evalDatasetsUsedAtLeastThreeTimes }{ 14 }
\newcommand{ \evalMetricUsageTopFour }{ 59.1 }

\newcommand{ \evalMetricUsagePointBased }{ 89.8 }
\newcommand{ \evalConfusionMatrixPapers }{ 59 }
\newcommand{ \evalConfusionMatrixPapersPecent }{ 9.7 }
\newcommand{ \evalOneOfTopFourMetrics }{ 354 }
\newcommand{ \evalAllOfTopFourMetrics }{ 82 }
\newcommand{ \evalAllConfusionAndTopFourMetrics }{ 20 }
\newcommand{ \evalAvgMetricsLatestYear }{ 3.4 }
\newcommand{ \evalRecallInAllPublications }{ 43.5 }

\newcommand{ \evalRecallInMorrisGasPublications }{ 79.2 }
\newcommand{ \evalWithCode }{ 21 }
\newcommand{ \evalSharePublicDatasetsRecentYear }{ 54.7 }

\newcommand{ \evalUniqueAuthorsRecent }{ 1109 }
\newcommand{ \evalFirstPublication }{ 2003 }

\newcommand{ \evalPubsLatestYear }{ 130 }
\newcommand{ \evalRecentRelativeYearlyIncrease }{ 40.9 }

\newcommand{ \evalPubsHistoricYears }{ 28 }

\newcommand{ \evalUniqueMetricsBeforeAggregation }{ 186 }
\newcommand{ \evalPapersWithOnlyDescriptions }{ 157 }
\newcommand{ \evalAvgMetricsPerPaper }{ 2.5 }
\newcommand{ \evalRecentMinRelativeYearlyIncreaseTopTenConferences }{ 7.2 }
\newcommand{ \evalRecentMaxRelativeYearlyIncreaseTopTenConferences }{ 25.5 }
\newcommand{ \evalAverageTopFiveCiteCount }{ 46.6 }
\newcommand{\hlnum}[1]{#1}
\usepackage{amssymb}
\usepackage{url}

\usepackage{flushend}
\usepackage{orcidlink}

\begin{document}

\date{}

\title{SoK: Evaluations in Industrial Intrusion Detection Research}

\author{
  {\textsc{Olav Lamberts$^*$}}\orcidlink{0009-0000-2502-2001}\\
  RWTH Aachen University\\
  lamberts@comsys.rwth-aachen.de
  \and
  {\textsc{Konrad Wolsing$^*$}}\orcidlink{0000-0002-7571-0555}\\
  Fraunhofer FKIE \\
  RWTH Aachen University \\
  wolsing@comsys.rwth-aachen.de
  \and
  {\textsc{Eric Wagner}}\orcidlink{0000-0003-3211-1015}\\
  Fraunhofer FKIE \\
  RWTH Aachen University \\
  eric.wagner@fkie.fraunhofer.de
  \and
  {\textsc{Jan Pennekamp}}\orcidlink{0000-0003-0398-6904}\\
  RWTH Aachen University \\
  pennekamp@comsys.rwth-aachen.de
  \and
  {\textsc{Jan Bauer}}\orcidlink{0000-0001-7631-6456}\\
  Fraunhofer FKIE \\
  jan.bauer@fkie.fraunhofer.de
  \and
  {\textsc{Klaus Wehrle}}\orcidlink{0000-0001-7252-4186}\\
  RWTH Aachen University \\
  wehrle@comsys.rwth-aachen.de
  \and
  {\textsc{Martin Henze}}\orcidlink{0000-0001-8717-2523}\\
  RWTH Aachen University \\
  Fraunhofer FKIE \\
  henze@spice.rwth-aachen.de
}

\maketitle

\def\thefootnote{*}\footnotetext{These authors contributed equally to this work.}
\def\thefootnote{}\footnotetext{This paper has been accepted at the Journal of Systems Research (JSys) Volume 3(1) 2023 \url{http://doi.org/10.5070/SR33162445}.}

\thispagestyle{empty}

\begin{abstract}
Industrial systems are increasingly threatened by cyberattacks with potentially disastrous consequences.
To counter such attacks, industrial intrusion detection systems strive to timely uncover even the most sophisticated breaches.
Due to its criticality for society, this fast-growing field attracts researchers from diverse backgrounds, resulting in \num{\evalPubsLatestYear} new detection approaches in 2021 alone.
This huge momentum facilitates the exploration of diverse promising paths but likewise risks fragmenting the research landscape and burying promising progress.
Consequently, it needs sound and comprehensible evaluations to mitigate this risk and catalyze efforts into sustainable scientific progress with real-world applicability.
In this paper, we therefore systematically analyze the evaluation methodologies of this field to understand the current state of industrial intrusion detection research.
Our analysis of \num{\evalAcceptedPaper} publications shows that the rapid growth of this research field has positive and negative consequences.
While we observe an increased use of public datasets, publications still only evaluate \num{\evalAverageDatasetsPerPaper} datasets on average, and frequently used benchmarking metrics are ambiguous.
At the same time, the adoption of newly developed benchmarking metrics sees little advancement.
Finally, our systematic analysis enables us to provide actionable recommendations for all actors involved and thus bring the entire research field forward.
\end{abstract}

\section{Introduction}
\label{sec:intro}

The digitalization of \acp{ICS} has led to an escalating rise in cyberattacks~\cite{Alladietal2020Industrial,Milleretal2021Looking,Sunetal2021SoK:}, of which prominent ones include the Stuxnet or Ukrainian power grid attacks.
These attacks are boosted by widely deployed legacy devices not meant to implement crucial security measures~\cite{Dahlmannsetal2022Missed}.
Specialized \acp{IIDS} address this gap by providing an easily retrofittable security solution for legacy industrial deployments~\cite{Dingetal2018A,Giraldoetal2018A}.
To this end, \acp{IIDS} passively monitor network traffic or the physical process state and alert human operators to initiate adequate countermeasures in case of suspected attacks~\cite{Wolsingetal2022IPAL:}.

As an emerging hot research area, \acp{IIDS} attract researchers and industrial operators from diverse backgrounds.
It thus comes as no surprise that, according to our literature research, at least \hlnum{\num{\evalUniqueAuthorsRecent}} distinct authors have published ideas for detection mechanisms between 2019 and 2021 alone.
While their diverse background is beneficial to cover lots of different perspectives and ideas, the resulting fast-paced advancements lead to a lack of established evaluation methodologies and comparability across the field.
Consequently, worthwhile ideas remain hard to identify, and it is unclear which improvements are suitable to close the gap to much-needed production-ready \acp{IIDS}.
E.g., specific detection methodologies have been independently proposed and evaluated for different \ac{ICS} domains in the past resulting in work being made twice~\cite{Wolsingetal2022IPAL:}.
While reproducibility studies are worthwhile, in contrast, accidental duplication of results likely leads to less significant overall research output.
Ideally, the vast research efforts would be channeled through clear, comparable, coherent, and expressive evaluation methodologies, that are updated upon new measurement discoveries.
Only through a resulting comparability between approaches, optimally performed for several \ac{ICS} domains, can the \ac{IIDS} research landscape fully benefit from its high diversity.

Digging deeper into conducted evaluations, researchers use benchmarking datasets that are either publicly available or, more commonly, custom-made for a specific evaluation and not published afterward, hindering repeatable experiments.
Based on these datasets and an \ac{IIDS}' alerts, various (performance) metrics are computed.
However, \acp{IIDS} are often evaluated on pre-selected datasets, covering specific favorable scenarios~\cite{Contietal2021A}.
Furthermore, metrics are chosen or designed based on specific goals determined (to some degree arbitrarily) by the researchers.
The resulting custom evaluation methodologies lead to an immense heterogeneity within the \ac{IIDS} research landscape, where most works, despite common goals, lack comparability.
Consequently, technological and scientific progress is inhibited.

In this regard, meta-analyses of \ac{IIDS} research already unveiled inefficiencies in the detection capabilities of published works~\cite{Erbaetal2020No} or criticized the conclusions drawn from scientific evaluation procedures~\cite{Kusetal2022A,Arpetal2022Dos,Fungetal2022Perspectives}.
Simultaneously, we observe attempts to fix these issues by, \eg collecting representative benchmarking datasets~\cite{Contietal2021A}, inventing specialized industrial metrics to accurately assess the ``success'' of an \ac{IIDS}~\cite{Huetetal2022Local,Kimetal2022A,Hwangetal2019Time-Series,Hwangetal2022Do,Gargetal2022An,Jamesetal2020Novel,Kimetal2022Towards,Lavinetal2015Evaluating,Tatbuletal2018Precision}, or by providing an abstract format to facilitate a coherent research landscape~\cite{Wolsingetal2022IPAL:}.
However, related work so far still fails to
\begin{enumerate*}[label=(\roman*)]
	\item quantify how \acp{IIDS} are evaluated within the vast body of literature,
	\item assess the applicability and impact of recent critiques partially known from, \eg traditional intrusion detection~\cite{ Sommeretal2010Outside,Milenkoskietal2015Evaluating,Arpetal2022Dos}, and
	\item deliver overarching recommendations to pave the way towards the shared goal of improving \acp{IIDS} to truly protect industrial networks and critical infrastructure against future cyberattacks.
\end{enumerate*}

With this paper, we strive to close the outlined gap with a \ac{SoK} on the evaluation methodologies across \ac{IIDS} research.
To this end, we conduct a \ac{SMS}, which is a variation of a systematic literature research more suited to obtain a broad view on a topic along potentially multiple research questions~\cite{Kitchenham2007Guidelines}.
With this methodology, we quantify the current state of the research landscape encompassing \hlnum{\num{\evalAcceptedPaper}} papers.
From the resulting knowledge basis, we can draw a clear picture \wrt positive and negative developments as well as persistent flaws.
Ultimately, our works allow us to provide clear recommendations for all involved actors to catalyze their joint efforts to protect the world's most critical networks.

\textbf{Contributions.}
To pave the way toward a more coherent \ac{IIDS} landscape, we make the following contributions:

\begin{itemize}[label=\textbullet,leftmargin=3mm]
	\item We survey \hlnum{\num{\evalAcceptedPaper}} papers published until 2021 	proposing \ac{IIDS} designs and extract information about how their respective evaluations were conducted (\secref{sec:survey}).
	\item We systematize the gained knowledge \wrt utilized datasets and metrics to identify positive and negative trends as well as their potential for future improvements.
	We then complement these \emph{theoretical} results with \emph{practical} experiments to extend the understanding of the interplay between datasets and metrics (\secref{sec:eval} and \secref{sec:metrics}).
	\item Finally, we summarize current flaws in \ac{IIDS} evaluations and formulate recommendations to improve future \ac{IIDS} research for all involved actors: \ac{IIDS} researchers, dataset creators, and industrial operators (\secref{sec:recommendations}).
\end{itemize}

\textbf{Artifact Availability.}
We make the data of our \ac{SMS} available at \url{https://zenodo.org/record/10008180}, and publish our evaluation tools used for the practical experiments at \url{https://github.com/fkie-cad/ipal_evaluate}.

\section{Research on Industrial Intrusion Detection}
\label{sec:bg}

To lay the foundation for our work, we provide a brief introduction to the general field of industrial intrusion detection (\secref{sec:bg:iids}) and present two \acp{IIDS} from literature in more detail in a case-study (\secref{sec:bg:casestudy}).
We then derive the challenges in \ac{IIDS} research (\secref{sec:bg:bg}) before we discuss related work on the evaluation methodologies of this research field (\secref{sec:bg:rw}).
Based on this, we motivate the need for systematizing the knowledge on evaluating industrial intrusion detection research and formulate basic research questions (\secref{sec:bg:need}) to ultimately steer future research in an effective direction.

\subsection{Industrial Intrusion Detection}
\label{sec:bg:iids}

The high degree of digitization in industries unleashes an enormous level of automation by integrating sensors, actuators, and control logic into tightly coupled cyber-physical systems.
The current trend to build \acp{ICS} by adapting once proprietary and local network protocols, \eg Modbus, to ubiquitous Ethernet networks, \eg using ModbusTCP, paired with connectivity to the Internet, enables unique applications, \eg remote monitoring or a more widespread deployment of \ac{SCADA} systems.
Yet, these technologies simultaneously open new attack vectors, as prominent attacks demonstrate~\cite{Alladietal2020Industrial,Milleretal2021Looking}.

To counter these security issues, various preventive measures have been proposed, \eg secure variants of industrial communication protocols~\cite{Contietal2021A,Dahlmannsetal2022Missed}.
However, deploying these protocols is difficult in practice, primarily due to industrial equipment's long lifetimes (up to 30 years)~\cite{Serroretal2021Challenges}.
These lifetimes arise to save cost and maximize availability but mean that deployed hardware is often hardly updateable and furthermore only has very limited processing power, such that deploying cryptographic protocols is often not feasible~\cite{Serroretal2021Challenges}.
But even for modern deployments, it is not always trivial to integrate secure communication protocols as they add significant bandwidth overhead and processing delays to a network that is already stressed due to the large number of devices in modern industrial networks~\cite{Serroretal2021Challenges}.
Less intrusive methods to retrofit preventive security measures through additional devices that authenticate traffic between them~\cite{Tsangetal2008YASIR:} or the embedding of authentication data directly in communications~\cite{Wagneretal2023Retrofitting} have also had little impact in practice, as commercial plug-and-play solutions are hardly available.

\begin{figure}
	\includegraphics[width=\columnwidth]{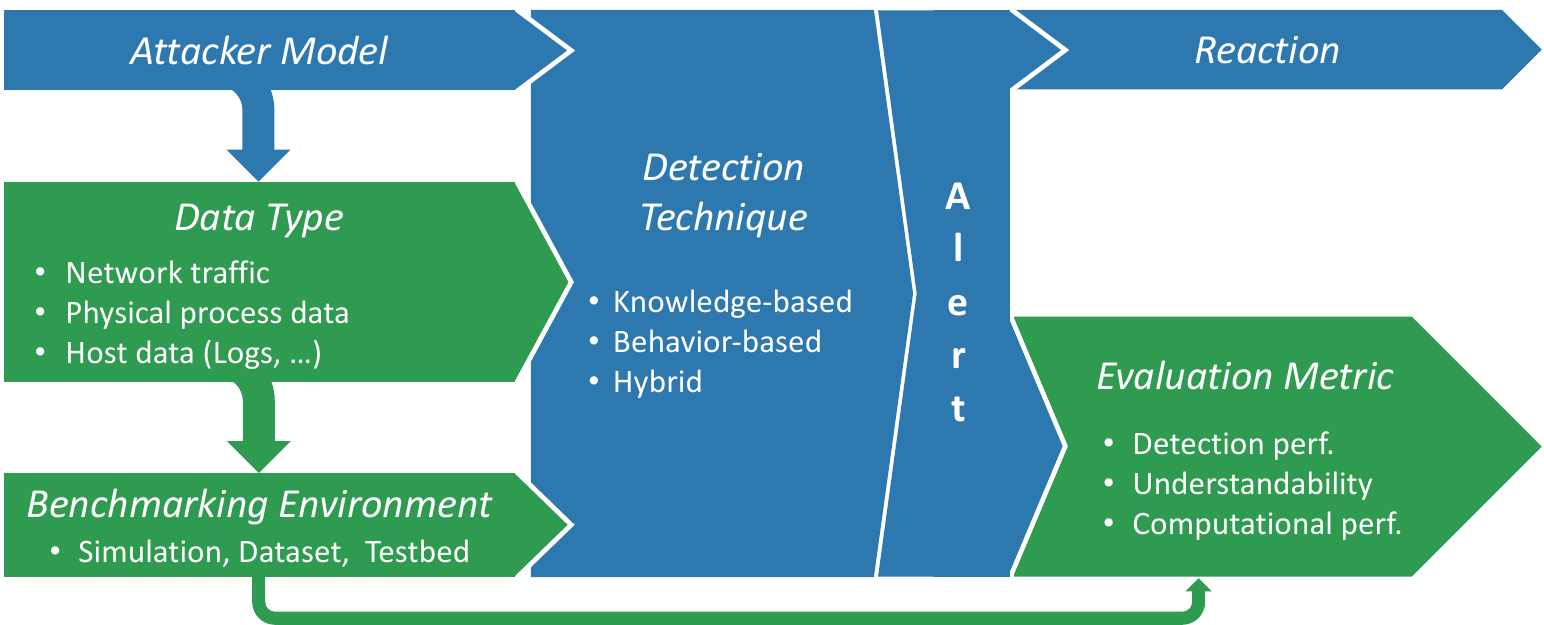}

	\caption{The industrial intrusion detection research domain can be roughly assessed along six essential dimensions. This \ac{SoK} tackles the associated dimensions to evaluate and compare \ac{IIDS} proposals (\cf dimensions marked in green).}
	\label{fig:iids:dimensions}
\end{figure}

In this context, intrusion detection is proposed as a promising alternative or complementing technology to passively retrofit security into \acp{ICS}~\cite{Wolsingetal2022IPAL:} by monitoring systems or networks for suspicious activities or violations of security policies.
However, established intrusion detection solutions from computer networks serving, \eg offices or data centers, are not as effective in industries~\cite{Zhouetal2015Design}, primarily due to \acp{ICS}' reliance on unique (real-time) hardware such as \acp{PLC} and sophisticated, custom-tailored attacks targeting the physical process~\cite{Urbinaetal2016Limiting,Alladietal2020Industrial}.
Consequently, research focuses on specialized \acfp{IIDS}, which leverage the repetitive and predictable characteristics occurring in, \eg Modbus' communication patterns or the physical process.
Overall, \ac{IIDS} research can be coarsely classified along six dimensions: attacker model, data type, detection technique, benchmarking environment, evaluation metric, and reactions (\cf \figref{fig:iids:dimensions}).

The \emph{attacker model} influences which kind of attacks an intrusion detection system should be able to detect and potentially even differentiate.
E.g., while a \acs{DoS} attack within the network can inhibit the real-time communication between \acp{PLC}, a malware reprogramming a \ac{PLC} can change the physical process.
For an extensive list of possible attack models, please refer to the MITRE ATT\&CK classification for \ac{ICS}~\cite{Alexanderetal2020MITRE}.
Note that while some surveys consider fault detection similar to attacks~\cite{Luoetal2021Deep}, faults do not occur as a consequence of cyberattacks but rather through natural processes and mistakes, \eg wear and tear~\cite{Giraldoetal2018A} and are thus left out of the scope of this work.
Thus, the attacker model greatly determines which \emph{data type} the \ac{IIDS} observes, with common ones being network traffic, \eg to detect a \ac{DoS} attack, host data from \ac{SCADA} systems or \acp{PLC}, \eg to detect unwanted modifications to the devices, and physical process data, \eg to find out whether the entire \ac{ICS} operates as intended~\cite{Luoetal2021Deep}.

The main work of researchers then goes into designing the actual \emph{detection technique}, which can be loosely categorized into knowledge-based, behavior-based, or hybrid approaches~\cite{Mitchelletal2014A}.
While knowledge-based systems (also referred to as misuse or supervised detection~\cite{Mitchelletal2014A}) identify harmful actions based on (known) patterns of cyberattacks, behavior-based \acp{IIDS} rather specify how the \ac{ICS} behaves normally and alert deviations from usual actions.
Moreover, the detection technique is also heavily influenced by the attacker model and data type.
While attacks on a network layer can be detected on a per-packet basis, \eg with deep-packet inspection~\cite{Goldenbergetal2013Accurate}, process-based detection leverages a broader view of an \ac{ICS}'s physical state, \eg by analyzing whether the process moves towards a critical state that may lead to an incident~\cite{Carcanoetal2011A}.
In \secref{sec:bg:casestudy}, we examine these two directions in more detail along two exemplary \acp{IIDS} from literature.

To validate the design of a detection technique and facilitate comparability of a newly proposed \ac{IIDS}, its detection performance is evaluated with the help of suitable \textit{benchmarking environments} and \emph{evaluation metrics} (potentially in addition to computational performance or \wrt explainability~\cite{Sengetal2022Why}).
Despite the data type, benchmarking environments for all kinds of industrial domains come in different forms, such as datasets, physical testbeds, simulations, or real facilities\cite{Kavallieratosetal2019Towards,Contietal2021A}.
Each type has its own trade-offs in terms of, \eg accessibility, cost, or closeness to real deployments, so their selection needs to be carefully made.
Moreover, the \ac{IIDS}' performance needs to be measured based on metrics enabling scientific comparisons between different proposals.
In that regard, scientists can refer to a plethora of common metrics~\cite{Powers2011Evaluation:}, such as precision, recall, the F1 score, or expressing the amount of false positive alerts, as well as even more complex characteristics (\cf \appdxref{sec:appx:metrics}).

A final dimension are \emph{reactions} to \ac{IIDS} alerts to mitigate an attack.
Especially when transferring an \ac{IIDS} to real-world deployments, operators may conduct (manual) forensic analyses to understand the cause for alert~\cite{Al-abassietal2022A} and ultimately mitigate the threat~\cite{Sunetal2021SoK:} by, \eg applying firewall rules.
Preventive measures can also be coupled to a detection mechanism for more automated reactions, then called intrusion prevention systems.
Those do, however, need to be carefully designed, since in an industrial setting simply blocking suspicious traffic may cause more harm than the suspected attack itself.

\subsection{Case Study on IIDSs}
\label{sec:bg:casestudy}

Having laid out the motivation and scope of industrial intrusion detection, the following section presents a case-study on two \ac{IIDS} proposals from literature and especially their underlying detection techniques in greater detail.
These two approaches represent diverging \ac{IIDS} directions:
The first approach analyzes network traffic with a knowledge-based detection model (\secref{sec:bg:casestudy:network}).
In contrast, the second \ac{IIDS} operates on the physical data of an \ac{ICS} leveraging behavior-based learning (\secref{sec:bg:casestudy:process}).
In the end, we shortly highlight the main differences of these directions (\secref{sec:bg:casestudy:comparison}).

\subsubsection{Knowledge-based, Network Intrusion Detection}
\label{sec:bg:casestudy:network}

The first case-study concerns a knowledge-based network \ac{IIDS}.
Here, we consider a minimalistic \ac{ICS} scenario where a \ac{PLC} controls a sensor and an actuator with ModbusTCP, \ie the \ac{PLC} requests a new measurement from the sensor and upon reception of the response commands the actuator to change its setpoint appropriately.
In this scenario, an attacker can ingest malformed packets, \eg to disrupt the actuator's behavior, since ModbusTCP is not authenticated by default~\cite{Dahlmannsetal2022Missed}.
The goal of an network \ac{IIDS} would be analyze the packets and indicate those that seem suspicious.

Two \acp{IIDS} implementing this direction are compared against each other by Perez \etal~\cite{Perezetal2018Machine}.
They leverage knowledge-based machine-learning models, \ie \acp{RF} or a \ac{SVM}, to classify observed network packets.
To this end, several features are extracted from each ModbusTCP packet such as the source, destination, packet length, or commanded setpoint values to the actuator.
Given a labeled training dataset containing exemplary benign \emph{and} malicious network packets, the \ac{RF} can learn patterns which indicate attacks by deriving suitable decision trees.
Finally, the trained model is used for live detection and can indicate on a per-packet basis whether a packet resembles an attack or is benign.

\subsubsection{Behavior-based, Process Intrusion Detection}
\label{sec:bg:casestudy:process}

Contrary to the previous example, the second case-study concerns behavior-based \acp{IIDS} operating, instead of network packets, on (system-wide) \ac{ICS} process data.
As \ac{ICS} scenario, we consider a facility with a water tank that has to maintain a minimum and maximum water level with an inlet valve filling the tank.
An attack against this system could result in a state where the valve is open even though the maximum water threshold is reached leading to an overflow.
The goal of a process-based \ac{IIDS} is to indicate these unwanted states.

One approach tackling this problem is presented by Feng \etal~\cite{Fengetal2019A}.
Their idea is to systematically find invariants that must be fulfilled all the time.
I.e., regarding the example an equation stating that if the inlet valve is open then we expect the water level to be below the maximum.
Since this rule is always satisfied during normal behavior of the \ac{ICS}, any violation resembles an unwanted situation.
Consequently, given a large training set of \emph{benign} \ac{ICS} behavior, this \ac{IIDS} can mine universally valid invariants and thus detect deviations without having to rely on sample attacks during training.

\subsubsection{Comparison}
\label{sec:bg:casestudy:comparison}

As can be seen from the two showcased \acp{IIDS}, intrusion detection can take vastly different forms in the \ac{ICS} domain.

First, while network traffic-based approaches detect attacks on a per-packet basis, \eg with deep-packet inspection, process-based IIDSs leverage a broader view of the ICS, \eg by analyzing the entire physical process data.
Consequently, both directions can detect widely different cyberattacks as network traffic-based \acp{IIDS} rather indicate \ac{MitM} attacks or malicious access to devices and process-based \acp{IIDS} detect the consequence of an attack to the physical process irrespective of the attack vector.
This difference is mainly influenced by the data type an \ac{IIDS} uses.

Independent of the data type, the underlying detection methodologies either train \emph{only} on benign samples to derive a normality model and indicate deviations (behavior-based, \cf \secref{sec:bg:casestudy:process}) or learn a combination of benign \emph{and} attack samples (knowledge-based, \cf \secref{sec:bg:casestudy:network}).
In that regard, one disadvantage of knowledge-based \acp{IIDS} is that they demand the \ac{ICS} to be under attack for training, which may be difficult to record especially in safety-critical \ac{ICS} deployments~\cite{Baderetal2023Comprehensively}.

While we shed light on these two dimensions, \acp{IIDS} can fall in between~\cite{Umeretal2022Machine}, \ie making use of knowledge-based and behavior-based detection methodologies.
This goes to show that not all \acp{IIDS} strictly follow the classification presented here and we refer the reader to other surveys which highlight the differences in greater depth~\cite{Luoetal2021Deep,Mitchelletal2014A,Umeretal2022Machine,Wolsingetal2022IPAL:}.

\subsection{Challenges of Evaluating IIDS}
\label{sec:bg:bg}

\ac{IIDS} research takes place in a diverse field encompassing \ac{ICS} architectures ranging from water supply over power delivery to manufacturing, where cyberattacks are primarily unique to a particular deployment~\cite{Alladietal2020Industrial,Milleretal2021Looking}.
Even though \acp{ICS} rely on researchers to design appropriate countermeasures and test their efficiency in real-world deployments, operators rarely provide such urgently-needed data samples~\cite{Sommeretal2010Outside,Ahmedetal2020Challenges,M.-R.etal2021Machine}.
While these challenges constitute an opportunity to tackle \ac{IIDS} research from varying angles, transfer insights across industrial domains, and investigate their efficiency in real-world deployments, they likewise segregate the overall research landscape, resulting in isolated silos~\cite{Wolsingetal2022IPAL:}.
Consequently, sound scientific evaluations remain as the foundation to facilitate coherence and measure the overall progress of the research field.

However, due to influences from various fields and a generally high interest in \acp{IIDS}, so far no coherent evaluation methodology could be established and subsequently improved.
In practice, the path taken by most researchers to design and test their \acp{IIDS} relies on privately acquired and/or public (synthetic) datasets containing samples of benign traffic and/or physical process data as well as attack scenarios.
To evaluate their \acp{IIDS}, researchers first train (and configure) their \ac{IIDS} on samples of \emph{benign} behavior and/or \emph{attacks} (depending on the type of \ac{IIDS}) from a specific industrial scenario.
On a second evaluation dataset, they then compare the \ac{IIDS} output~(alerts) to the attack labels contained within the chosen dataset, \ie they track how well the \ac{IIDS} detects attacks and to which degree benign traffic or process values are unintentionally classified as suspicious.
Finally, various metrics, \eg the F1 score, quantify the detection performance and serve as the basis for comparisons to related work.

While most works adhere to this loosely outlined evaluation methodology, the devil is in the details~\cite{Kusetal2022A}.
Optimally, a given dataset would be suitable for a large amount of \ac{IIDS} types and thus constitute a reference benchmark.
However, widely-used datasets usually cover only specific industrial domains and a small subset of imaginable attacks~\cite{Contietal2021A}.
Thus, the datasets made available to the research community decisively influence the scenarios within which \acp{IIDS} are evaluated and also the types of attacks \acp{IIDS} are optimized for.
Moreover, utilized evaluation metrics do not draw a complete picture of an \ac{IIDS}'s detection performance without putting them into context~\cite{Giraldoetal2018A}, which rarely happens adequately within the research field.
As a matter of fact, this lack of hardened and proven research methodologies has been exposed to various criticism in recent years, as identified by related work.

\subsection{Related Work on Evaluating IIDSs}
\label{sec:bg:rw}

Taking a closer look at recent literature on the challenges of evaluating industrial intrusion detection research (\cf \secref{sec:bg:bg}), we identify a range of works discussing and criticizing the current state of \ac{IIDS} research.
First, various surveys provide an overview of the utilized \emph{detection methods} across that research field~\cite{Sunetal2021SoK:,Luoetal2021Deep,Mitchelletal2014A,Zhangetal2021A,Dingetal2018A,Giraldoetal2018A,Umeretal2022Machine,Wolsingetal2022IPAL:,Rakasetal2020A}, ranging from learning specific communication patterns to analyzing the physical state of the monitored system.
In this context, difficulties reproducing results and generalizing \acp{IIDS} to related \ac{ICS} domains beyond those specifically evaluated were reported~\cite{Erbaetal2020No,Wolsingetal2022IPAL:}.
While these surveys repeatedly cover more than \num{70}~publications, showing the huge attention industrial intrusion detection attracts, at the same time, they indicate a lack of coherence and advancement within the research field.

Similar surveys focused on summarizing available \emph{datasets} and testbeds~(from which datasets can be generated) specifically designed for \ac{IIDS} evaluations~\cite{Kavallieratosetal2019Towards,Contietal2021A}.
These efforts identify at least \num{61} testbeds and \num{23} benchmarking datasets that are publicly available~\cite{Contietal2021A}.
Since these surveys focus solely on datasets, they lack essential analyses about the actual application of datasets.
As a rare exception, Balla \etal~\cite{Ballaetal2022Applications} analyzed dataset usage for deep learning detection methodologies, observing a strong bias toward non-\ac{ICS} datasets, such as the KDD dataset family representing a collection of datasets to evaluate \acp{IDS} mostly from the computer network domain~\cite{Dhanabaletal2015A}, with a usage of over \SI{50}{\percent}.

Besides datasets or testbeds, the choice of \emph{metrics} is important when evaluating \acp{IIDS}.
Without a dedicated focus on \emph{industrial} intrusion detection, Powers~\cite{Powers2011Evaluation:} provides an overview of different metrics and puts their expressiveness into context.
Yet, the considered point-based metrics (\cf \appdxref{sec:appx:metrics:pbm}), \eg accuracy or precision (also used in other domains such as machine learning), must be used carefully not to introduce any biases~\cite{Powers2011Evaluation:}.
In that regard, Christen\etal~\cite{Christenetal2024A} performed a in-depth analysis of the F1 score across many research disciplines.
For evaluations on (industrial) time-series datasets, further challenges, such as an imbalanced representation of attacks, have to be considered~\cite{Gensleretal2014Novel,Arpetal2022Dos}.
Thus, more advanced time series-aware metrics have been proposed~\cite{Huetetal2022Local,Kimetal2022A,Hwangetal2019Time-Series,Hwangetal2022Do,Gargetal2022An,Jamesetal2020Novel,Lavinetal2015Evaluating,Tatbuletal2018Precision}~(\cf \appdxref{app:time-series-aware-metrics}).
While this development promises to enhance the expressiveness of evaluations, their soundness and usage remain mostly unexplored.

Finally, various \emph{meta-surveys} focus on machine learning pitfalls for industrial intrusion detection~\cite{Rakasetal2020A,Etalle2017From,M.-R.etal2021Machine,Fungetal2022Perspectives,Lanferetal2023Leveraging,Umeretal2022Machine,Turrinetal2020A,Apruzzeseetal2023SoK:} or highlight challenges when transferring \acp{IIDS} from research to actual industrial deployments~\cite{Sommeretal2010Outside,Ahmedetal2020Challenges,M.-R.etal2021Machine}.
These problems include, \eg inappropriate use of metrics~\cite{Arpetal2022Dos}, the dominance of lab-based datasets~\cite{Arpetal2022Dos,Rakasetal2020A}, analysis of dataset quality~\cite{Lanferetal2023Leveraging,Turrinetal2020A}, or predominant focus on only a few of the wide range of industrial domains and protocols~\cite{Rakasetal2020A}.
W.r.t. network-based intrusion detection leveraging machine learning, not just for \acp{ICS}, Apruzzese \etal~\cite{Apruzzeseetal2023SoK:} propose a new concept of pragmatic assessments that advocates evaluating more toward the values perceived by industrial operators actually implementing \acp{IDS} in the end.
Importantly, empirical data on the evaluation of \acp{IIDS} is not yet available.

In summary, evaluations of \acp{IIDS} can, in theory, be based on a solid foundation of public datasets and advanced metrics.
However, this research branch lacks a decent understanding of the methodologies actually applied within it beyond individual criticism regarding isolated aspects.

\subsection{The Need for Systematization}
\label{sec:bg:need}

The tremendous research interest in industrial intrusion detection, with \hlnum{\num{\evalPubsLatestYear}} publications in the year 2021 alone, has led to a huge variety of evaluation methodologies.
The resulting fast-paced research has a huge risk of becoming disjoint~\cite{Wolsingetal2022IPAL:}, eventually slowing down the overall progress in securing \acp{ICS}.
Most importantly, the heterogeneity across industrial domains~\cite{Wolsingetal2022IPAL:} and an observed widespread evaluation bias~\cite{Giraldoetal2018A,Urbinaetal2016Limiting,Wolsingetal2022IPAL:} make comparisons between \acp{IIDS} difficult.
Past surveys on detection methodologies, datasets, metrics, and meta-studies have only studied individual aspects in isolation from each other (cf. \secref{sec:bg:rw}).
Thus, to unveil the root causes hindering coherent and sustainable \ac{IIDS} research, there is a need to systematically consolidate the current state of evaluations in industrial intrusion detection research to ultimately identify remedies against the status quo.

We argue that only by analyzing how \acp{IIDS} are evaluated on a broad scale, as done in a \acf{SMS}~\cite{Kitchenham2007Guidelines}, we can comprehensively tackle the question of research coherence and evaluation soundness, \ie to which extent evaluations are performed on uniform (public) datasets with widespread and suitable metrics to achieve a high level of comparability.
More precisely, we aim to answer the following research questions:

\noindent $\blacktriangleright$ \textbf{Q1:} Which datasets are actually used to evaluate \acp{IIDS}? \\
\noindent $\blacktriangleright$ \textbf{Q2:} Which metrics are utilized in evaluations of \acp{IIDS}? \\
\noindent $\blacktriangleright$ \textbf{Q3:} To which extent do \acp{IIDS} compare against each other?

Besides providing a comprehensive picture of the traits and characteristics of \ac{IIDS} evaluations, answering these questions lays the foundation to formulate actionable recommendations for \ac{IIDS} evaluation, enabling the different actors within the research community to focus their joint effort on the overarching challenge of securing industrial deployments.

\section{Systematic Mapping Study}
\label{sec:survey}

The objective of this \ac{SoK} is to provide a systematic understanding of how (differently) \ac{IIDS} research is currently evaluated and how this current status quo can be sustainably improved.
While related work already hints at prevalent issues that might prevent objective comparisons (\cf \secref{sec:bg:rw}), a holistic analysis is missing so far.

Therefore, we strive to ascertain the state of \ac{IIDS} evaluation methodologies by conducting a \acf{SMS}, a variation of a classical \ac{SLR}~\cite{Kitchenham2007Guidelines}, to obtain a large, qualitative, and unbiased collection of relevant publications in a verifiable process, oriented along established best practices and guidelines~\cite{Kitchenham2007Guidelines}.
In contrast to a systematic literature review, a \ac{SMS} addresses a broader scope usually indicated by formulating multiple research questions and, as a consequence, results in a more general search string with many hits which can imply a slightly vague data extraction process due to the diversity of found publications~\cite{Kitchenham2007Guidelines}.
In return, its broader scope enables a \ac{SMS} to identify further research topics which demand detailed attention from future research.

\begin{figure}[t]
	\centering
	\includegraphics[width=\columnwidth]{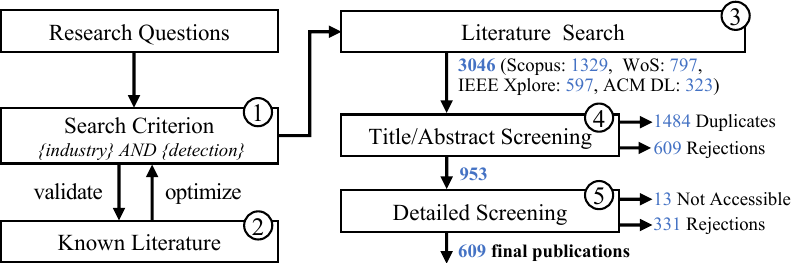}
	\caption{To conduct the \ac{SMS}, we follow a two-staged approach which results in extracting a total of \hlnum{\num{\evalAcceptedPaper}} relevant publications proposing novel \ac{IIDS} as of December~2022.
	We list the corresponding search string in \appdxref{sec:appx:slr}.}
	\label{fig:survey-design}
\end{figure}

To holistically answer the outlined research questions for a large and heterogeneous research field, we perform a comprehensive \ac{SMS} as depicted in \figref{fig:survey-design}.
To this end, we first search relevant papers for a broad subject (\ac{IIDS} proposals) from the scientific literature with a systematic process.
Afterward, publications are analyzed and classified based on the subjects of our analysis (Q1--Q3), \ie their evaluation methodology.
According to the research questions, the \ac{SMS} focuses on publications that propose \acp{IIDS} for \acp{ICS} as researchers naturally have to evaluate their performance in a scientific manner.
In contrast to Balla \etal~\cite{Ballaetal2022Applications}, we only consider publications that leverage at least one \textit{industrial}-specific dataset, \ie they were obtained from an \ac{ICS}, \eg include specific protocols such as Modbus, physical process data, or \ac{ICS}-specific cyberattacks.

To conduct our \ac{SMS}, we leverage Parsifal~\cite{Freitas2013Parsifal} to organize and comprehensibly document our screening process.
First, we transformed the research questions into a search string~\circled{1} (\cf \appdxref{sec:appx:slr}), which we successively optimized through validation with an initial set of known and representative literature~\circled{2}.
We then queried four search engines~(IEEE Xplore, ACM DL, Scopus, and Web of Science) on December 2022 and found a total of \hlnum{\num{3046}} hits~\circled{3}.
From this initial set of publications, we discarded duplicates (\hlnum{1484} publications) and performed a first screening of all remaining publications' titles and abstracts~\circled{4}.
In this initial screening, we mostly focused on removing publications from other research domains that still matched our search string and such publications that clearly do not propose~(and thus evaluate) an \ac{IIDS} approach.
After this first screening phase, \hlnum{\num{953}} unique publications remained.
Note that we did not filter for any specific detection techniques.
Still, most publications covered by the survey (and thus the research field) resemble machine-learning.

In a final step, we conducted a detailed screening of the remaining publications to extract those that build the foundation for our further analysis~\circled{5}.
When accessing the full text of all papers, only \hlnum{\num{13}} publications were not accessible to us and thus omitted.
We performed a detailed second screening of all remaining publications, resulting in \hlnum{\num{331}} further rejections of those that do not match our requirements for proposing \acp{IIDS}, \eg belonged to fault detection (\cf \secref{sec:bg:iids}).
From the resulting set of \hlnum{\num{\evalAcceptedPaper}} accepted publications, we extracted the relevant data to answer our research questions, such as the datasets and metrics they utilize for their evaluations.
To ensure consistency, one author performed the detailed screening and data extraction while the workload for initial title/abstract screening was shared across multiple persons.

Through our systematic approach, to the best of our knowledge, we are the first to analyze the entire \ac{IIDS} landscape.
With \num{\evalAcceptedPaper} analyzed publications, our work is based on a significantly larger knowledge base than any of the previous surveys of related work (\cf \secref{sec:bg:rw}).
This basis enables us to analyze the evaluation methodologies of the broad \ac{IIDS} research landscape.
Beyond presenting our findings, releasing our \ac{SMS} as a public artifact (\cf Artifact Availability) may help future researchers to find appropriate candidates for comparisons, facilitates further analyses, or enables tracking of the progress within the \ac{ICS} domain in the future.

\section{Evolution of IIDS Research and Datasets}
\label{sec:eval}

With a systematic basis of \hlnum{\num{\evalAcceptedPaper}} publications proposing \acp{IIDS} gathered in our \ac{SMS} (\cf \secref{sec:survey}), we initially assess how the overall research landscape on \acp{IIDS} has evolved over time.
As a systematic representation has been missing so far~(\cf \secref{sec:bg:rw}), we augment the field with a high-level overview in \secref{sec:eval:overview}.
Afterward, we unveil common trends in evaluation methodologies, especially \wrt the utilized datasets (\cf \secref{sec:eval:datasets}) and represent the two most common representatives in a case-study in more detail (\cf \secref{sec:eval:dataset:casestudy}).

\subsection{Overview of the IIDS Research Landscape}
\label{sec:eval:overview}

We begin our analysis with a high-level overview of the evolution and composition of the \ac{IIDS} research landscape.

\begin{figure}
	\includegraphics[width=\columnwidth]{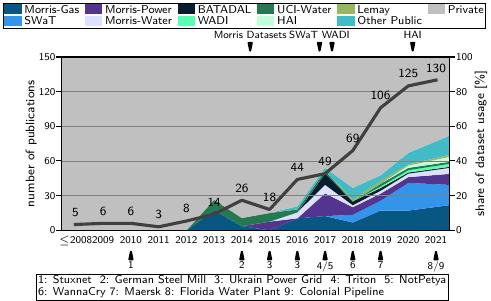}

	\caption{Publications on \acp{IIDS} took off around 2013 and kept increasing as more cyberattacks occurred. Simultaneously a trend fosters to evaluate on public datasets.}
	\label{fig:pubs-over-time}
\end{figure}

\subsubsection{Evolution}
To understand the evolution of the \ac{IIDS} research domain, we focus on the number of published papers over time (\cf \figref{fig:pubs-over-time}), which we enrich with timestamps of notable cyber
incidents and the releases of commonly used evaluation datasets.
While the first publications within the \ac{IIDS} domain date back to \hlnum{\num{\evalFirstPublication}}, the domain initially received little attention, with only \hlnum{\num{\evalPubsHistoricYears}} publications until 2012.
From 2013 onward, research took off exponentially, with an average increase of \hlnum{\SI{\evalRecentRelativeYearlyIncrease}{\percent}} in yearly publications.
In 2021, the last year considered in our \ac{SMS}, we identified \hlnum{\evalPubsLatestYear} new publications, which is higher than in any previous year.
In comparison, the Top 10 cyber security conferences experienced a lower average yearly increase in accepted publications from \hlnum{\SI{\evalRecentMinRelativeYearlyIncreaseTopTenConferences}{\percent}} for Crypto up to only \hlnum{\SI{\evalRecentMaxRelativeYearlyIncreaseTopTenConferences}{\percent}} for USENIX Sec during the same timespan~\cite{Zhou2009Top}.

We presume that the key driver for the interest in this research domain is caused by the raised public awareness following the Stuxnet cyberattack and subsequent ones like the two incidents with the Ukrainian power grid~\cite{Alladietal2020Industrial}.
Apart from such targeted attacks, industries were equally affected by more widespread malware, such as NotPetya or WannaCry~\cite{Alladietal2020Industrial}, due to their increasing digitalization and Internet-facing deployments (\cf \secref{sec:bg:iids}).
With attacks continuing~\cite{Milleretal2021Looking}, endangering human safety, expensive equipment, and the environment, the peak in 2021 with \hlnum{\num{\evalPubsLatestYear}} proposals comes as no surprise---underlining the growing relevance of \ac{IIDS} research.

A first look at the (publicly) utilized datasets' in \figref{fig:pubs-over-time} also allows us to deduce the existence of a growing number of public datasets.
These datasets stem from various industrial domains, such as water purification, gas distribution, and electrical power generation, among many others.
This conclusion aligns with recent results identifying a growing number of public datasets emerging across many industrial domains~\cite{Contietal2021A}.

\textbf{Takeaway.}
From this assessment, we conclude that \ac{IIDS} research increasingly tackles the diverseness of industrial domains based on variously utilized datasets and experiences steady growth that does not seem to have reached its peak yet.

\subsubsection{Coherence}
\label{sec:eval:overview:coherence}

\begin{figure}
	\centering
	\includegraphics[width=\columnwidth]{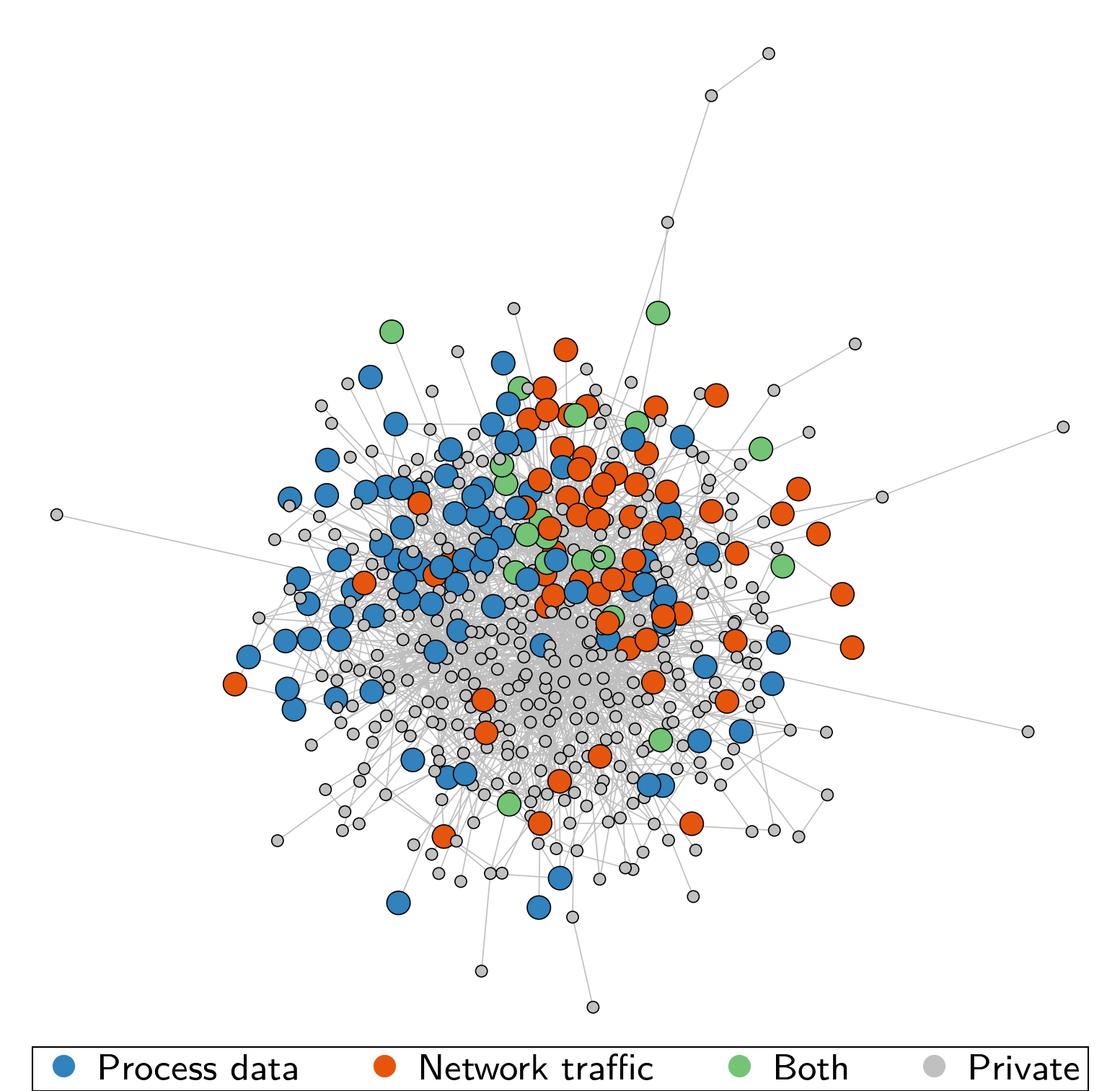}

	\caption{Publications arranged in a citation graph reveal two directions roughly disjunct into approaches considering network traffic datasets and ones evaluating process data.}
	\label{fig:citation-graph}
\end{figure}

For such a rapidly growing research landscape in a diverse industrial environment, we further want to understand how coherent research is performed, \ie whether the domain is split in loosely connected topics or which directions receive the most attention.
Therefore, we visualize the connections among publications by their \emph{citation relationships} in \figref{fig:citation-graph}.

Citation data was retrieved and aggregated from OpenAlex and Semantic Scholar for all \hlnum{\num{\evalAcceptedPaper}} publications, and we draw a connection between two publications if one cites another.
In \figref{fig:citation-graph}, publications are arranged by the Kamada-Kawai force-directed placement algorithm~\cite{Kamadaetal1989An}.
Moreover, for publications utilizing publicly accessible datasets, we colored their vertices belonging to process data datasets, network traffic, or both.
Note, however, that our analysis omits \hlnum{\num{\evalCitationGraphOmitted}} publications for which no connection to other publications could be found, either because the citation data for the respective publications was incomplete or because the \acp{IIDS} were indeed presented without relating to the vast body of existing works.

On average, a publication is cited by \hlnum{\num{\evalAverageCiteCount}} other \ac{IIDS} publications, while the Top \hlnum{\num{5}} cited publications~\cite{Carcanoetal2011A,Urbinaetal2016Limiting,Goldenbergetal2013Accurate,Hadziosmanovicetal2014Through,Kravchiketal2018Detecting} (not in order) are cited on average by \hlnum{\num{\evalAverageTopFiveCiteCount}} papers as of the 1st March 2023.
These numbers provide a first glance at the connectivity in \ac{IIDS} research.
However, note that this observation neither implies that a publication builds on top of prior works from the cited publication nor that they compare to that publication within their evaluations.

While inspecting the citation structure, we observe that the \ac{IIDS} research domain is divided into two basic directions based on the evaluated dataset types:
A first group of \hlnum{\num{\evalProcessDatasetPapers}} papers (blue) resembles the larger class that focuses on process data datasets.
A second slightly smaller class of \hlnum{\num{\evalNetworkDatasetPapers}} publications (red) corresponds to intrusion detection methodologies detecting attacks in network data.
Only rarely (\hlnum{\num{\evalBothDatasetPapers}} times) does a publication contain evaluations for both classes (green).
Interestingly, both research areas show little connectivity, indicating a limited exchange of knowledge across these fields.
This is backed by the fact that the clustering coefficient for the sub-domains (process data \hlnum{\num{\evalProcessClusteringCoefficient}} and network traffic \hlnum{\num{\evalNetworkClusteringCoefficient}}) is slightly higher than for the entire \ac{IIDS} landscape (\hlnum{\num{\evalGlobalClusteringCoefficient}}).

\textbf{Takeaway.}
Publications are more likely to cite each other if they stem from the same type, which promises a high number of comparisons among them.
Still, the low clustering indicates incoherence in the overall research domain.

\subsection{Benchmarking Datasets}
\label{sec:eval:datasets}

With a basic understanding of the \ac{IIDS} research domain, we now assess how evaluations are conducted in more detail.
In this context, the chosen benchmarking dataset is a crucial building block as it serves as the basis for nearly all subsequent performance calculations.
While related work has assessed which datasets are readily available~\cite{Contietal2021A}, their exact usage and distribution remains unknown as of now (\cf~\secref{sec:bg:rw}).
Consequently, this section answers our first research question Q1, regarding the datasets \acp{IIDS} are evaluated on.

\subsubsection{Overview}
\label{sec:eval:dataset:overview}

As can be derived from \figref{fig:pubs-over-time}, over the entire timespan, the majority of used datasets are private, and only \hlnum{\SI{\evalPapersWithPublicDatasetPercent}{\percent}} of the publications evaluate at least one public dataset.
Note that we counted datasets as private if there existed no obvious procedure to retrieve the dataset.
While private datasets may represent unique use cases, \eg real-world data of industrial facilities, they significantly hinder reproducibility and comparisons to related works since they usually deny access to outsiders.
In our \ac{SMS}, we refrained from investigating private datasets in more depth because of the varying degrees of descriptions throughout the publications.
Hence, needed details cannot be fully captured or verified.
Nonetheless, we observe a trend starting around 2013 toward increased utilization of public datasets, which accounts for \hlnum{\SI{\evalSharePublicDatasetsRecentYear}{\percent}} of the evaluated datasets in 2021.
Therefore, it is more likely that an \ac{IIDS} uses public datasets if published recently.

This trend follows the publication of high-quality datasets that are still widely used today.
When looking at peak usage of public datasets, the SWaT~\cite{Gohetal2016A} and Morris-Gas Pipeline~\cite{Morrisetal2015Industrial} datasets jointly occur in \hlnum{\SI{\evalSWaTMorrisGasPercentage}{\percent}} of the publications, which is the majority of the publications utilizing a public dataset at all (\hlnum{\SI{\evalPapersWithPublicDatasetPercent}{\percent}}) and other public datasets are thus used much less frequently.
As a consequence, a significant portion of research activities seems to be biased toward these two datasets.

Regarding dataset diversity, across our entire \ac{SMS}, we identified \hlnum{\num{\evalUniquePublicDatasets}} unique public datasets, which exceeds previous reports of \num{23} datasets by Conti \etal~\cite{Contietal2021A}.
In contrast to Balla \etal~\cite{Ballaetal2022Applications} (\cf \secref{sec:bg:rw}) and by the design of our \ac{SMS} (\cf \secref{sec:survey}), we dominantly encounter specialized industrial datasets contradicting their observed research bias toward non-industrial datasets.
However, of the many public datasets, \hlnum{\num{\evalDatasetsUsedOnce}} are only used once, and \hlnum{\num{\evalDatasetsUsedAtLeastThreeTimes}} occur at least three times (the Top nine public datasets are depicted in \figref{fig:pubs-over-time}).
Thus, availability alone is not decisive for a widespread use and other factors such as covered domain and attacks as well as the overall quality of the data seems to play an essential role as well.

\subsubsection{Case Study on Evaluation Datasets}
\label{sec:eval:dataset:casestudy}

To establish a more profound understanding of which forms \ac{ICS} datasets can take, we explore the two most common ones occurring in \hlnum{\SI{\evalSWaTMorrisGasPercentage}{\percent}} of all publications in a case-study:
Morris-Gas~\cite{Morrisetal2015Industrial} resembling a network capture mostly leveraged to evaluate knowledge-base \ac{IIDS} approaches and SWaT~\cite{Gohetal2016A} containing \ac{ICS} process data.
To this end, we first provide a brief overview and then discuss their differences \wrt \ac{IIDS} evaluations.
For a detailed description, please refer to their documentation or the survey conducted by Conti \etal~\cite{Contietal2021A}.

\textbf{Morris-Gas}~\cite{Morrisetal2015Industrial} represents a miniature gas-pipeline consisting of a pump, a solenoid release valve, and a pressure sensor, which should maintain a constant pressure in the system.
To this end, these devices are connected with Modbus to a central \ac{PLC}, which implements monitoring and control logic.
In addition, a \ac{HMI} serves as a display to monitor the process.
Then, the authors implemented $35$ attacks against Modbus, divided into seven categories that target the communication, \eg via \ac{DoS}, or the process itself, \eg by changing the target pressure value.
The final dataset consists of \num{214.580} Modbus packets, of which \num{60.048} are randomly chosen attacks.
Note that attacks are executed several times, sometimes even with slight variations in parameterization.
Each datapoint resembles a single Modbus packet containing as features generic Modbus elements, such as a packet's function ID, and process-level information, \eg{} the pressure measurements included in the payload, among others.
Most importantly, all packets are labeled individually for \ac{IIDS} training, whether they resemble the original communication or are caused by one of the attack types.

\textbf{Secure Water Treatment (SWaT)}~\cite{Gohetal2016A} was recorded on a scaled-down physical testbed of a real water treatment plant containing six stages:
water intake, chemical assessment and dosing, filtration, dechlorination, inorganic purification, and disposal.
Each stage is controlled by a \ac{PLC}, which also manages the synchronization between the stages, and the whole process is monitored in a \ac{SCADA} system.
In contrast to the Morris-Gas dataset, SWaT's process is much more complex, involving more than \num{50} sensors and actuators, depending on the dataset version, and has a stronger focus on process data than network communication.
Therefore, the main dataset is a collection of the physical state encompassing all current sensor and actuator readings every second, accompanied by optional network traffic.
For the most common dataset version (A1\&A2), 36 different cyberattacks were launched as \ac{MitM} between any stage's \ac{PLC} and the \ac{SCADA} system.
Here, attacks solely resemble physical modifications of the system, \eg shutting down a pump.
During the dataset recording, the testbed ran for eleven days in total, of which the first seven contain only normal operating behavior, and the latter four include the attacks for \ac{IIDS} testing.
Note that SWaT's attacks were not repeated and can last up to several hours.

\textbf{Comparison.}
When comparing the datasets' properties, it becomes apparent that they target different \ac{IIDS} types.
First and foremost, they significantly differ \wrt the scale and domain of the testbed, as well as, the provided data type.

Besides that, the composition of training and testing data is an equally important evaluation factor.
In that regard, SWaT's low number of attacks without repetitions renders it nearly useless to analyze an \acp{IIDS}' capability to detect one precise attack class, whereas Morris-Gas repeatedly executed the same attacks.
Vice versa, Morris-Gas does not provide a dedicated set of benign-only data, which is necessary for behavior-based approaches.

Another major factor affecting subsequent evaluations is the duration of an individual attack.
As a consequence of SWaT's large testbed and slow process cycle, it was reported that it could take some time after a cyberattack was initiated until effects become visible, and these effects can exceed the actual attack until the process stabilizes again, making determining the range of attacks difficult for evaluations~\cite{Turrinetal2020A}.
In contrast, the Morris-Gas dataset expects the \ac{IIDS} to precisely indicate each faulty network packet right at the time of observation.
Thus, the datasets' labels differ strongly in interpretation.
While for the Morris-Gas dataset, it is crucial to detect any network packet, for SWaT, one can argue that detecting an attack sometime during execution suffices.
Such effects strongly affect the interpretation of evaluation metrics, as we also discuss later (\cf \secref{sec:metric:experiment}).

\begin{table}
	\centering
	\setlength{\tabcolsep}{1pt}

	\begin{tabular}{cl Hc llr}

		\textbf{Origin} & \textbf{Name} & \textbf{Public} & \textbf{Type} & \textbf{Domain} & \textbf{Protocol} & \textbf{Usage} \\ \toprule
	
		\multirow{3}{*}{iTRUST$^a$}
		& SWaT~\cite{Gohetal2016A} & \good & P$^*$ & Water & -- & \hlnum{\SI{\evalSWaTUsedInPaperPercent}{\percent}} \\
		& BATADAL~\cite{Taorminaetal2018Battle} & \good & P\phantom{$^*$} & Water & -- & \hlnum{\SI{\evalBATADALUsedInPaperPercent}{\percent}} \\
		& WADI~\cite{Ahmedetal2017WADI:} & \good & P\phantom{$^*$} & Water & -- & \hlnum{\SI{\evalWADIUsedInPaperPercent}{\percent}} \\ \midrule
	
		\multirow{3}{*}{Morris et al.$^b$}
		& Morris-Gas~\cite{Morrisetal2015Industrial} & \good & N\phantom{$^*$} & Gas & Modbus & \hlnum{\SI{\evalMorrisGasUsedInPaperPercent}{\percent}} \\
		& Morris-Power~\cite{Adhikarietal2013Industrial} & \good & P\phantom{$^*$} & Electricity & -- & \hlnum{\SI{\evalMorrisPowerUsedInPaperPercent}{\percent}} \\
		& Morris-Water~\cite{Morrisetal2015Industrial} & \good & N\phantom{$^*$} & Water & Modbus & \hlnum{\SI{\evalMorrisWaterUsedInPaperPercent}{\percent}} \\ \midrule
	
		\multirow{4}{*}{Misc}
		& UCI-Water~\cite{Poch1993Water} & \good & P\phantom{$^*$} & Water & -- & \hlnum{\SI{\evalWaterPlantUsedInPaperPercent}{\percent}} \\
		& HAI~\cite{Shinetal2020HAI} & \good & P\phantom{$^*$} & Diverse & -- & \hlnum{\SI{\evalHAIUsedInPaperPercent}{\percent}} \\
		& Lemay~\cite{Lemayetal2016Providing} & \good & N\phantom{$^*$} & Electricity & Modbus & \hlnum{\SI{\evalLemayUsedInPaperPercent}{\percent}} \\
		\bottomrule
	\end{tabular} \\[.2em]
	N: Network captures \quad P: Process data\\[.5mm]
	$^*$ Network captures for SWaT exist, but are rarely used in research. \\[-.5mm]
	$^a$ \url{https://itrust.sutd.edu.sg/itrust-labs_datasets} \\[-.5mm]
	$^b$ \url{https://sites.google.com/a/uah.edu/tommy-morris-uah/ics-data-sets}

	\caption{Across the top nine public datasets, two account for the majority of uses. Despite \acp{ICS}' diversity, the top datasets focus on a few domains and protocol combinations.}
	\label{tab:datasets-overview}
\end{table}

\begin{figure}
	\centering
	\includegraphics[width=\columnwidth]{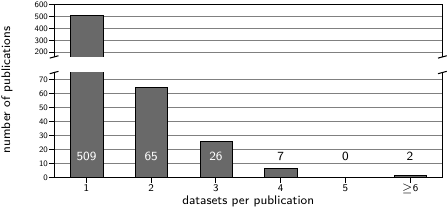}

	\caption{Publications usually utilize a single dataset, and only \hlnum{\SI{\evalPercentagePaperWithMultipleDatasets}{\percent}} of the papers leverage multiple datasets at all.}
	\label{fig:eval:datasets-per-publication}
\end{figure}

\subsubsection{Dataset Types}
\label{sec:eval:dataset:types}

In the next step, we examine the Top nine datasets more closely and highlight their different directions (\cf \tabref{tab:datasets-overview}).

First, a dataset's data \emph{type} can be either a network capture, mostly required for network-based \acp{IIDS} or a (preprocessed) sample of physical system data, \eg a time series of temperature values.
For each type, we observe one major origin that accounts for most of the utilization across research, with iTRUST for process-based datasets and Morris \etal primarily for network-based ones.
Considering the type of the top nine utilized datasets, we observe a strong focus on process-based datasets with \hlnum{\SI{20.3}{\percent}} compared to \hlnum{\SI{15.6}{\percent}} for network-based, which is in line with the observations from \secref{sec:eval:overview:coherence}.

Since industrial domains are diverse, we expect a large coverage of them across utilized datasets as well.
However, the commonly covered industrial domains are mainly driven by the water and gas facilities, indicating an underrepresentation of all other domains, such as power generation, electricity distribution, or manufacturing.
Yet, considering the large numbers of domains covered by private datasets, for which (high-quality) public alternatives do not exist, we cannot conclude that other domains receive few attention nor that those industries show no interest in \ac{IIDS} research.

Lastly, industries are well known for their diverse and incompatible pooling of network protocols, mostly for legacy reasons~\cite{Dahlmannsetal2022Missed}.
Despite market-share studies identifying \num{11} dominant network technologies~\cite{HMS-Networks2021Continued}, research either focus on Modbus (having \SI{10}{\percent} market share~\cite{HMS-Networks2021Continued}) or no communication protocol at all.
While we discovered \acp{IIDS} for further industrial protocols such as IEC 60870-5-104~\cite{Linetal2019Timing}, S7~\cite{Linetal2017Timing-Based}, or DNP3~\cite{Radoglou-Grammatikisetal2020DIDEROT:}, their representation is marginal and mostly confined to private datasets.
Therefore, the distributions of utilized datasets \wrt their type, industrial domain, and network protocol reveal a significant drift between peer-reviewed literature and actual production systems.

\begin{table}
	\centering

	\begin{tabular}{l c c}
		\textbf{Combination} & \textbf{Count} & \textbf{Origins} \\ \toprule
		Morris-Gas \& Morris-Water & \hlnum{12} & 1 \\
		Morris-Gas \& Morris-Power & \hlnum{8} & 1 \\
		Morris-Power \& Morris-Water & \hlnum{7} & 1 \\
		SWaT \& WADI & \hlnum{4} & 1\\
		Morris-Gas \& UCI-Water & \hlnum{4} & 2 \\
		Morris-Gas \& SWaT & \hlnum{3} & 2 \\
		Electra Modbus \& S7Comm & \hlnum{3} & 1 \\
		\midrule
		Morris-Gas, Power \& Water & \hlnum{5} & 1 \\
		\bottomrule
	\end{tabular} \\[.2em]
	No private datasets were considered

	\caption{If multiple datasets are used, they mostly stem from the same class or origin, attributing little to richer evaluations.}
	\label{tab:dataset-combinations}
\end{table}

\subsubsection{Research Embedding}

In the last step, we assess how the different datasets are embedded into research.
Therefore we begin with the number of different datasets that are used within a single publication, as shown in \figref{fig:eval:datasets-per-publication}.
A large class of publications~(\hlnum{\num{\evalPaperWithOneDataset}}) evaluates a single dataset, and only a minority (\hlnum{\num{\evalPaperWithMultipleDatasets}}) on more than one.
One publication uses \hlnum{\num{\evalAverageDatasetsPerPaper}} datasets on average.
This observation is in line with the previous clustering observed in \secref{sec:eval:overview:coherence}, which is more coherent \wrt the top-used datasets, suggesting that researchers often primarily focus on a single dataset.
Given that we found at least \hlnum{\num{\evalUniquePublicDatasets}} datasets publicly available, researchers most likely could consider additional, compatible datasets, especially when claiming that proposed \acp{IIDS} are applicable to a large range of industrial domains~\cite{Wolsingetal2022IPAL:}.
This claim is backed by the fact that two publications have already evaluated as many as six datasets~\cite{Baptisteetal2021Systematic,Gianietal2014Phasor}.
However, our results also suggest a discrepancy between datasets \wrt ease of use, documentation, and completeness, motivating the limited use of the available datasets.

Looking into the preferred datasets, \tabref{tab:dataset-combinations} enumerates the top dataset combinations.
While we observe prominent combinations, the corresponding datasets usually originate from the same source and thus represent similar domains and protocols.
Only seven publications evaluate datasets that stem from two origins.
Thus, potentially widely applicable \acp{IIDS} are evaluated for specific (research) deployments from a single industrial domain, most likely not representative of an entire domain.
Consequently, research fails to effectively widen the scope of available evaluations and rather introduces biases by focusing on a few specific niches.

\textbf{Takeaway.}
\ac{IIDS} research is still governed by private datasets, with an increasing trend toward public datasets.
However, we observe a potential for improvement in the number of datasets used during evaluation as well as their diversity \wrt type, industrial domain, and network protocol.

\section{Survey on Evaluation Metrics}
\label{sec:metrics}

Previously, we analyzed the \ac{IIDS} research landscape \wrt its historic evolution and utilized datasets.
However, our analysis still lacks a detailed view at the leveraged evaluation metrics as they resemble an essential element of scientific evaluations (\cf \figref{fig:iids:dimensions}).
Moreover, and most importantly, it is still unclear how expressive a given (combination of) metric(s) is in judging the detection performance of an \ac{IIDS}.

To this end, we first provide an overview of common and newly proposed metrics and categorize them into a taxonomy~(\cf \secref{sec:metric:taxonomy}).
Afterward, we showcase how the two \acp{IIDS} from the case-study leveraged the chosen datasets and metrics to conduct their evaluations (\cf \secref{sec:metric:casestudy}).
Next, we take a broader view and assess the utilization of metrics across \ac{IIDS} research along our \ac{SMS}~(\cf \secref{sec:metrics:utilization}).
Having analyzed the leveraged datasets and metrics, we can ultimately assess the degree of reproducibility and comparability in \ac{IIDS} reserach (\cf \secref{sec:eval:comparisons}).
Finally, because of reports about flaws to established metrics, as indicated in \secref{sec:bg:rw}, we examine how susceptible the research domain is in that regard by analyzing their expressiveness in practical experiments~(\cf \secref{sec:metric:experiment}).

\subsection{A Taxonomy of IIDS Evaluation Metrics}
\label{sec:metric:taxonomy}

\begin{table}[t]
	\centering
	\setlength{\tabcolsep}{.5pt}
	\renewcommand{\arraystretch}{1.2}

	\definecolor{accuracy}{HTML}{3182bd}
	\definecolor{precision}{HTML}{6baed6}
	\definecolor{recall}{HTML}{9ecae1}
	\definecolor{fone}{HTML}{c6dbef}
	\definecolor{confusion}{HTML}{e6550d}
	\definecolor{fpr}{HTML}{fd8d3c}
	\definecolor{tnr}{HTML}{fdae6b}
	\definecolor{fnr}{HTML}{fdd0a2}
	\definecolor{curvebased}{HTML}{74c476}
	\definecolor{detectiondelay}{HTML}{756bb1}
	\definecolor{timeaware}{HTML}{9e9ac8}
	\definecolor{npv}{HTML}{ffc0cb}
	\newcommand{\tikzcircle}[1][red,fill=red]{\tikz[baseline=-0.5ex]\draw[fill=#1,radius=2pt] (0,0) circle ;}

	\begin{tabular}{cl ccccHH lHH}
		&
		\textbf{Metric \hfill Appendix} &
		\multicolumn{4}{c}{\tikzcircle[confusion] \textbf{Conf.\ Matr.}} &
		Multi-class &
		\textbf{Time} &
		\quad\quad\quad\textbf{Synonyms} &
		\textbf{Usage} &
		\textbf{App.} \\
		\cline{3-6}

		& & TP & TN & FP & FN & MC & & \\ \toprule

		\multirow{11}*{\rotatebox{90}{\textbf{Point-based}}}
		& \multirow{2}*{\tikzcircle[recall] TPR \hspace{1.9cm} \ref{app:tpr}} & \multirow{2}*{\checkmark} & & & \multirow{2}*{\checkmark} & & & Recall, Sensitivity & & \multirow{2}*{\ref{app:tpr}} \\
		& & & & & & & & Hit-Rate & & \\
		& \tikzcircle[fnr] FNR \hfill \ref{app:fnr} & \checkmark & & & \checkmark & & & Miss-Rate & & \ref{app:fnr} \\
		& \tikzcircle[tnr] TNR \hfill \ref{app:tnr} & & \checkmark & \checkmark & & & & Specificity, Slectivity & & \ref{app:tnr} \\
		& \tikzcircle[fpr] FPR \hfill \ref{app:fpr} & & \checkmark & \checkmark & & & & Fall-out & & \ref{app:fpr} \\
		& \tikzcircle[precision] PPV \hfill \ref{app:ppv} & \checkmark & & \checkmark & & & & Precision, Confidence & & \ref{app:ppv} \\
		& \tikzcircle[npv] NPV \hfill \ref{app:npv} & & \checkmark & & \checkmark & & & -- & -- & \ref{app:npv} \\
		\cline{2-11}

		& \tikzcircle[accuracy] Accuracy \hfill \ref{app:acc} & \checkmark & \checkmark & \checkmark & \checkmark & & & Rand Index & & \ref{app:acc} \\
		& \tikzcircle[fone] F1 \hfill \ref{app:f1} & \checkmark & & \checkmark & \checkmark & & & -- & & \ref{app:f1} \\
		\cline{2-11}

		& \tikzcircle[curvebased] RoC \hfill \ref{app:roc} & \checkmark & \checkmark & \checkmark & \checkmark & & & -- & & \ref{app:roc} \\
		& \tikzcircle[curvebased] AuC \hfill \ref{app:auc} & \checkmark & \checkmark & \checkmark & \checkmark & & & -- & & \ref{app:auc} \\
		\cline{2-11}

		\hline

		\multirow{4}*{\rotatebox{90}{\textbf{\parbox{1.7cm}{Time-aware}}}}
		& \tikzcircle[timeaware] Detected Scenarios \hfill \ref{app:detscen} & \checkmark & & & & & \checkmark & -- & & \ref{app:detscen} \\
		& \tikzcircle[detectiondelay] Detection Delay \hfill \ref{app:detdelay} & \checkmark & & & \checkmark & & & -- & & \ref{app:detdelay} \\
		& \tikzcircle[timeaware] (e)TaPR~\cite{Hwangetal2022Do,Hwangetal2019Time-Series} \hfill \ref{app:tapr} & \checkmark & \checkmark & \checkmark & \checkmark & & & eTaP, eTaR, eTaF1 & & \ref{app:tapr} \\
		& \tikzcircle[timeaware] Affiliation~\cite{Huetetal2022Local} \hfill \ref{app:affiliation} & \checkmark & \checkmark & \checkmark & \checkmark  & & & -- & & \ref{app:affiliation} \\

		\bottomrule

	\end{tabular}

    \caption{Our taxonomy distinguishes between point-based and time series-aware metrics. Metrics may occur under different synonyms. For details, refer to \appdxref{sec:appx:metrics}.}
	\label{tab:metric:taxonomy}
\end{table}

Evaluating the performance of an \ac{IIDS} is of utmost importance to prove its effectiveness and compare it quantitatively against related works either in terms of attack detection performance, or computational resources.

Since \emph{computational} resources are stated only occasionally throughout the \ac{SMS}, we shorty introduce which aspects were evaluated.
The most prominent aspect, still in \num{136} publications, refers to the time to train a model or classify a given datapoint/dataset.
More infrequently are statistics about CPU/GPU usage (\num{13}), RAM utilization (\num{12}), or model size (\num{16}).
However, a sound comparison without equivalent hardware or implementations is challenging and therefore those metrics are beyond the scope of the \ac{SoK} in the following.

Regarding \emph{detection} performance, during the conduction of our \ac{SMS}, we extracted a total of \hlnum{\num{\evalUniqueMetricsBeforeAggregation}} distinct metrics that were used during the evaluations.
To provide an initial holistic overview, we present the most used metrics found in the \ac{SMS} and relevant (newer) ones observed in related work in a taxonomy (\cf \tabref{tab:metric:taxonomy}).
The metrics are discussed in a more general fashion in the following, while short explanations for all {\num{14} introduced metrics can be found in the \appdxref{sec:appx:metrics}.

\subsubsection{Confusion Matrix}
\label{sec:metric:taxonomy:confusion}

Scientific evaluations of \acp{IIDS} base on labeled benchmarking datasets (\cf \secref{sec:eval:datasets}), including samples of cyberattacks (malicious) and benign behavior.
After a training phase, for each data-point in the dataset, the known labels are compared to the output of the \ac{IIDS} (alarm or no alarm).
The high-level goal of an \ac{IIDS} is to detect as many attack instances as possible while emitting few (false) alarms for benign behavior.
Note that especially in \acp{ICS}, where cyberattacks are rare compared to benign behavior, false alarms should be minimal~\cite{Etalle2017From}.

As the first performance indicators, one can count the occurrences of all four possible combinations between dataset labels and \ac{IIDS} outcomes called true-positive (TP), true-negative~(TN), false-positive~(FN), and false-positive~(FP), making up the confusion matrix to capture an \ac{IIDS}s behavior.

\subsubsection{Point-based Metrics}

Since there is a desire to express performance with a single value irrespective of the dataset, there exist a large variety of \emph{point-based} metrics derived from the confusion matrix~\cite{Powers2011Evaluation:} (\cf \tabref{tab:metric:taxonomy}).
These express properties, such as its overall correctness (accuracy), the fraction of correct alarms (precision), or fraction of identified attacks (recall).
Point-based metrics find wide application beyond \ac{IIDS} research, \eg machine learning, and thus a natural choice for comparisons.

\subsubsection{Time Series-aware Metrics}
Point-based metrics are suitable when the benchmarking datasets' entries are independent.
However, \acp{ICS} are inherently time-dependent, \ie the current state of an \ac{ICS} is always a result of the system's previous state.
Consequently, \ac{IIDS} datasets extracted from these systems also need to be considered in the aspect of time, \ie an alarm extending beyond an attack while the system did not yet reach its normal operational state should be interpreted differently from a false alarm in the middle of normal behavior.
In such or similar scenarios, point-based metrics are skewed, which is already known in literature since 2014 by Gensler \etal~\cite{Gensleretal2014Novel}.

Consequently, many novel \emph{time-aware} metrics tackle such flaws~\cite{Huetetal2022Local,Hwangetal2022Do,Hwangetal2019Time-Series,Tatbuletal2018Precision,Lavinetal2015Evaluating,Gensleretal2014Novel}.
They, \eg simply count the number of detected and continuous attack scenarios (detected scenarios)~\cite{Linetal2018TABOR:}, aggregate the time it takes until the \ac{IIDS} emits an alert after the attack began (detection delay), or define new time series-aware versions of precision and recall to favor early detection of an attack instance (e)TaPR~\cite{Hwangetal2022Do}.
Yet, Huet \etal~\cite{Huetetal2022Local} already found that (e)TaPR is not free of flaws and responded with their own Affiliation metric.
Note that while time series-aware metrics like Numenta~\cite{Lavinetal2015Evaluating} or the one proposed by Tatbul \etal~\cite{Tatbuletal2018Precision} and Gensler \etal~\cite{Gensleretal2014Novel} exist, they were observed only seldom in our \ac{SMS}, if at all.

\subsection{Case-Study on IIDS evaluations}
\label{sec:metric:casestudy}

Prior to examining the metrics leveraged in the entire \ac{IIDS} research domain, we take a closer look at the two publications from our case-study and how they evaluated their approach.

Perez \etal~\cite{Perezetal2018Machine} evaluate the knowledge-based network \acp{IIDS} with the help of the Morris-Gas dataset~\cite{Morrisetal2015Industrial}.
Since their classifiers, \ie \ac{RF} and \ac{SVM}, require examples of benign \emph{and} malicious traffic for training, they shuffle the entire dataset randomly and split it into three parts for training, validation, and evaluation.
After having determined optimal hyperparameters for their detection models, they leverage the evaluation part to examine the \acp{IIDS}' performance.
To this end, they compare the classifiers output with the labels of the dataset and calculate the accuracy, precision, recall, and F1 score.
In addition to the overall scores, Perez \etal also provide a detailed confusion matrix of all attack types for the best classifier (\ac{RF}).
Note that cross-validation, as known from machine learning~\cite{Refaeilzadehetal2016Cross-Validation}, was not conducted by the authors.
Moreover, the methodology of randomly shuffling the dataset has the drawback that the model has likely seen samples of all attacks during training and thus risks detecting only those later on instead of detecting novel attacks~\cite{Kusetal2022A}.
In the end, Perez \etal do not compare their results to related work.

In contrast, Feng \etal evaluate their approach on two distinct datasets (SWaT and WADI)~\cite{Fengetal2019A}.
As both datasets already ship with two parts, one containing solely benign system behavior and one with attacks (\cf \secref{sec:eval:dataset:casestudy}), random shuffling or splitting the dataset is not necessary.
Moreover, this assures that the behavior-based \ac{IIDS} only trains on attack free data.
As evaluation metrics, Feng \etal calculate recall, fall-out, as well as the number of detected scenarios.
In addition, they provide a normalized version of recall which accounts for the different lengths of the attacks in the dataset.
Besides analyzing the performance of their \ac{IIDS}, Feng \etal also compare their results to two different approaches.

As apparent from these two examples, \ac{IIDS} approaches can be evaluated quite diversely, not only \wrt to the underlying dataset, as analyzed in detail in \secref{sec:eval:datasets} before, but also in terms of evaluation methodologies.
This diversity involves the experiment design and conduction, \eg performing a train/validation/test split or not requiring those at all.
Consequently, assessing the precise evaluation procedure or how the final metrics were calculated for such a broad class of publications is challenging in general.
To this end, we take a closer look at the metrics to measure \ac{IIDS} performance and their utilization in research in the following which can still provide valuable insights into the comparability among scientific publications and the overall expressiveness of evaluations.

\subsection{Metrics Utilized in IIDS Research}
\label{sec:metrics:utilization}

Given that a wide variety of metrics exist to express \ac{IIDS} performance, in our next research question, Q2, we ask how often and when these metrics are used.
Overall in our \acp{SMS}, we found \hlnum{\num{\evalUniqueMetricsBeforeAggregation}}
different metrics and flavors, including subtle deviations such as multi-class or weighted variants.
As an example, multi-class classification distinguishes between the different attack types.
Thus a metric, such as multi-class TPR, can provide greater insight into which attack type gets detected the best by a specific approach.
To handle this amount of metric and flavors in the following, we aggregated them into similar classes, \eg binary-class and multi-class TPR are simply considered and listed as TPR.
Note that we took special care to handle some edge cases during this aggregation, such that, \eg multi-class variants are mapped to their \emph{semantic} equivalent in the binary case.
This is partly motivated by the fact that most papers only use binary metrics.
E.g., if a publication states multi-class TPR also for the class of benign data class, we counted this publication for TPR and additionally for TNR, as the TPR over the benign class is equivalent to TNR.
Lastly, since a majority of the metrics is used infrequently, \ie only \hlnum{\num{\evalMetricsUsedAtLeastTenTimes}} occur at least ten times, we bundle rarer metrics into a single class (others).

\subsubsection{Metrics over time}
\label{sec:metrics:utilization:overtime}

\begin{figure}[t]
	\includegraphics[width=\columnwidth]{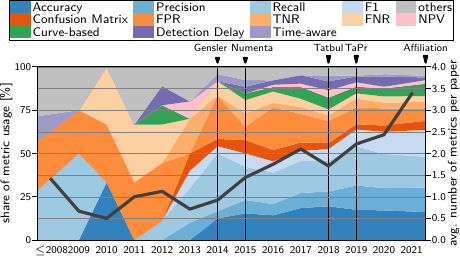}

	\caption{Point-based metrics dominate \ac{IIDS} evaluations, with accuracy, precision, recall, and F1 being the top most used metrics. Over time, the number of metrics in a publication increased to currently \hlnum{\num{\evalAvgMetricsLatestYear}} on average in 2021.}
	\label{fig:metric-over-time}
\end{figure}

To obtain a first overview of the utilization of frequent metrics, we depict their use over time in \figref{fig:metric-over-time}.
First of all, the number of different metrics used in a single publication on average (\hlnum{\num{\evalAvgMetricsPerPaper}} overall) kept increasing since 2013, and nowadays, publications use \hlnum{\num{\evalAvgMetricsLatestYear}} metrics on average.
This greatly coincides with the previous observation in \figref{fig:pubs-over-time}, where the year 2013 marked the turning point when \ac{IIDS} research took off.
This trend toward more metrics contributes to higher comparability in the research domain and hints at in-depth evaluations.
However, there also exist \hlnum{\num{\evalPapersWithOnlyDescriptions}} publications that evaluate without any quantitative metrics and instead rely only on textual descriptions, \eg elaborating which attack scenarios were detected or discussing results visually along graphs.
Note that textual descriptions cannot be aggregated into a unified class as they differ significantly, \ie two publications using textual descriptions hardly describe the same feature.

In contrast to dataset utilization (\cf \figref{fig:pubs-over-time}), the metric utilization fluctuates less over time.
One notable trend, again starting around 2013, is that accuracy, precision, recall, and F1, \ie the classical point-based metrics, have established themselves as metrics with high usage by representing \hlnum{\SI{\evalMetricUsageTopFour}{\percent}} of all used metrics.
At the same time, out of the \hlnum{\num{\evalOneOfTopFourMetrics}} publications utilizing one of these four metrics, only \hlnum{\num{\evalAllOfTopFourMetrics}} state all four.
Thus their usage is inconsistent, and most publications only focus on certain aspects of their expressiveness.

Concerning all point-based metrics, which account for \hlnum{\SI{\evalMetricUsagePointBased}{\percent}} of all metrics, the confusion matrix resembles an important metric as it builds the foundation to calculate all point-based metrics (\cf \secref{sec:metric:taxonomy:confusion}).
However, out of the \hlnum{\num{\evalConfusionMatrixPapers}} papers that publish the confusion matrix, just \hlnum{\num{\evalAllConfusionAndTopFourMetrics}} fully state or discuss all four common metrics (accuracy, precision, recall, and F1), even though this would be easily doable.
In the \hlnum{\SI{\evalConfusionMatrixPapersPecent}{\percent}} of publications where the confusion matrix is published, at least missing metrics can be calculated, which is not possible the other way round, \ie the confusion matrix cannot be computed if, \eg F1 scores are indicated.
It thus remains questionable why publications omit frequently used metrics when all data to compute them has to be available anyway.

Even though it has been known since 2014 that for industrial IDSs, point-based metrics may be flawed~\cite{Gensleretal2014Novel}, they make up \hlnum{\SI{\evalMetricUsagePointBased}{\percent}} of all metrics.
As a time series-aware metric, detection delay receives constant but infrequent use by \hlnum{\num{\evalPapersWithDetectionDelay}} publications overall.
Still, detection delay alone does not quantify the portion of detected attacks and thus likely serves to enhance point-based metrics.
Newer promising time series-aware metrics yet have to gain traction (only \hlnum{\num{\evalPapersWithTimeaware}} publications use them), despite their added value in interpreting \ac{IIDS} results.

\begin{figure}
	\includegraphics[width=\columnwidth]{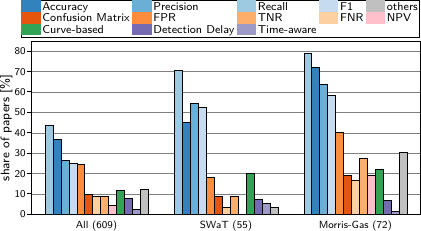}

	\caption{Papers utilizing SWaT and especially the Morris-Gas dataset are dominated by point-based metrics. Time series-aware metrics are slightly more frequent for SWaT.}
	\label{fig:metric:utilization}
\end{figure}

Evaluations in the \acp{IIDS} research dominantly build upon point-based metrics, which are known to have flaws, especially on time-series datasets as used in \ac{IIDS} research~\cite{Gensleretal2014Novel,Hwangetal2019Time-Series}.

\subsubsection{Metric distribution on datasets}
\label{sec:metrics:utilization:dataset}

Taking a closer look at the metric utilization, it may be interesting to investigate the differences between industrial domains or datasets since we already observed the formation of obvious clusters in research around publications using the same type of datasets in \secref{sec:eval:overview:coherence}.
However, as not all \ac{IIDS} publications clearly state which \ac{ICS} domain their approach is designed for, we opted to concentrate solely on the selected evaluation dataset.
Therefore we pick the two most commonly used datasets, SWaT and Morris-Gas, representing not only two different dataset types (process data and network captures) but also different domains (Water and Gas), as apparent from \tabref{tab:datasets-overview}.
Consequently, \figref{fig:metric:utilization} depicts the dataset's influence on the chosen metrics by comparing their metrics distribution against the set of all publications.

The top four metrics (accuracy, precision, recall, and F1) play a major role for the SWaT and Morris-Gas datasets too, even more than across all publications.
Recall, for example, is used in \hlnum{\SI{\evalRecallInAllPublications}{\percent}} of all publications but indicated for \hlnum{\SI{\evalRecallInMorrisGasPublications}{\percent}} of \acp{IIDS} evaluated on the Morris-Gas dataset.
The order of usage between them is also similar, \ie recall is used the most and F1 the least.
The only exception is accuracy, which is indicated less often for the SWaT dataset.
This difference might be caused by SWaT featuring far fewer attack instances.
Another exception is that other point-based metrics (confusion matrix, FPR, TNR, FNR, and NPV) receive greater attention in the Morris-Gas dataset.
Contrary, time series-aware metrics are slightly more common for SWaT.
We suspect that the favor for time-aware metrics in SWaT stems from the dataset nature consisting of longer-lasting attacks compared to individual malicious network packets in the Morris-Gas dataset (\cf \secref{sec:eval:dataset:casestudy}).

Our analysis highlights once again the dominance of point-based metrics, especially for the top two datasets by usage.

\subsubsection{Metric Combinations}
\label{sec:metrics:utilization:combinations}

\begin{figure}[t]
	\includegraphics[width=\columnwidth]{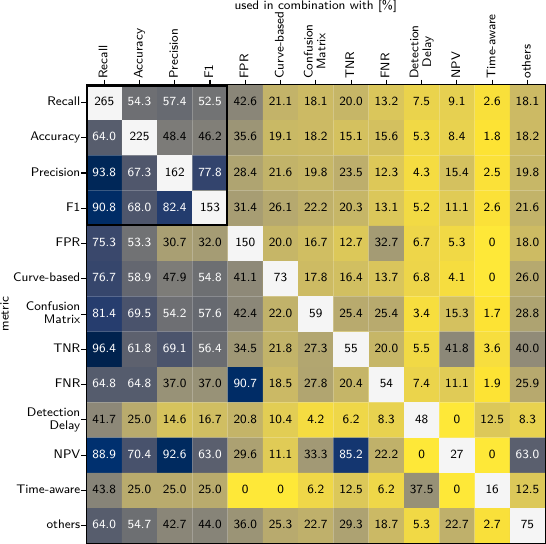}

	\caption{Metrics show strong correlations \wrt combinations they occur in publications. \Eg F1 is used in \hlnum{\num{\evalPapersWithFScore}} publications, and among them \hlnum{\SI{82.4}{\percent}} publish precision. Vice versa, \hlnum{\SI{77.8}{\percent}} of the \hlnum{\num{\evalPapersWithPrecision}} papers with precision also state F1.}
	\label{fig:metric:combinations}
\end{figure}

Even though there exists a variety of metrics (\cf \secref{sec:metric:taxonomy}), a single metric usually has to be considered in relation to others.
E.g., precision and recall have to be discussed jointly since an \ac{IIDS} which detects all attacks (high recall score) might do so simply by emitting alerts continuously, which would become visible in a low precision score.
Fused metrics like F1 try to remedy this situation but deny in-depth reasoning afterward as they do not retain the precise original information.
According to our \ac{SMS}, publications state \hlnum{\num{\evalAvgMetricsPerPaper}}
metrics on average to sketch light on the \ac{IIDS} performance from different perspectives.
Consequently, as the last step, we evaluate which metrics are used together.

To this end, \figref{fig:metric:combinations} depicts the occurrence of combinations between the considered metrics.
On the diagonal, we enumerate how often each metric is utilized globally, \ie recall is used \hlnum{\num{\evalPapersWithRecall}}
times.
The remaining cells indicate how often the indication of one metric leads to the usage of another metric.

In total, \hlnum{\num{\evalPapersWithFScore}}
publications used the F1, and \hlnum{\SI{90.8}{\percent}} of these papers (stating F1) also published recall values.
This is not surprising since knowledge of the recall is required to calculate F1.
Vice versa, however, \hlnum{\num{\evalPapersWithRecall}}
papers used recall, and of them only \hlnum{\SI{52.5}{\percent}} of those also published F1 scores.
Looking at precision and recall as two complementing metrics, recall is used in \hlnum{\SI{93.8}{\percent}} of the publications that state precision.
If recall is stated, only \hlnum{\SI{57.4}{\percent}} also publish precision.
While the number of detected attacks (recall) is valuable information, at least \hlnum{\SI{20.38}{\percent}} of papers using recall do not include \emph{any} measure, \eg accuracy, which depends on the \ac{IIDS} correctly classifying benign data-points, \ie TNs.
Note that this number is a lower bound since we could not always verify whether benign data was included, \eg in averages of multi-class metrics.

For popular point-based metrics (within the black rectangle), we observe a strong dependence between them, which is not surprising as these are heavily used (\cf \figref{fig:metric-over-time} and \figref{fig:metric:utilization}).
Since many point-based metrics are derived from the confusion matrix (\cf \tabref{tab:metric:taxonomy}), the confusion matrix likewise has a high correlation with these four.
However, it is not guaranteed that these are published reliably, as F1 is contained in only \hlnum{\SI{57.6}{\percent}} of the cases when the confusion matrix is presented.
This is in line with our previous observation that of the \hlnum{\num{\evalPapersWithConfusionMatrix}}
publications with a confusion matrix, only \hlnum{\num{\evalAllConfusionAndTopFourMetrics}} state all of the four most often used point-based metrics (\cf \secref{sec:metrics:utilization:overtime}).

Except for the dependencies between FNR and FPR, there exist few apparent correlations, thus often omitting the classical point-based metrics completely.
Especially publications taking advantage of newer, time series-aware metrics lack other metrics.
While this development makes sense (why should we indicate flawed metrics when we can use better ones), it makes comparisons to prior works harder.

\subsection{Reproducibility and Comparability}
\label{sec:eval:comparisons}

Finally, we address our third research question Q3 asking to which extent \acp{IIDS} compare against each other.
We assess this question from two directions, first by examining the conditions for reproducibility and second by measuring the degree of comparability, which are both perceived as good scientific standards~\cite{Peisertetal2007How}, even though reproducibility lacks far behind expectations in the entire research community (beyond intrusion detection research)~\cite{Baker20161500}.
While reproducibility enables researchers to comprehend, build upon, or even enhance existing work, comparability allows them to determine how well an approach performs, \ie to highlight the impact of newly proposed contributions over previous work or which approaches might be suitable for real-world deployments.

\subsubsection{Reproducibility}
\label{sec:eval:comparison:reproducibility}

Within \ac{IIDS} research, reproducing existing work is not uncommon, \eg to concisely analyze the prospects and limitations of individual approaches~\cite{Erbaetal2020No,Kusetal2022A}, prove the feasibility of new ideas upon reproduced implementations~\cite{Wolsingetal2022IPAL:}, or solely for scientific profoundness~\cite{Peisertetal2007How}.
Yet, successfully reproducing approaches is not guaranteed~\cite{Erbaetal2020No}.
To even enable the cumbersome process of reproducing \ac{IIDS} research, the availability of artifacts, such as datasets or code, is needed.

In our survey, we observe that \hlnum{\SI{\evalPapersWithPublicDatasetPercent}{\percent}} of the publications already utilize public datasets with an improving trend (\hlnum{\SI{\evalSharePublicDatasetsRecentYear}{\percent}} of utilized datasets in 2021 are public; \cf \figref{fig:pubs-over-time}).
However, successfully reproducing older publications is less likely.
While the availability of code is not strictly required, as the relevant details should be part of the publication, it greatly eases the reproducibility process.
Unfortunately, it is difficult to ascertain the availability of source code in a systematic way as it is not always clear where to find availability statements or corresponding pointers in publications.
Still, we only encountered \hlnum{\num{\evalWithCode}} publications with obvious references, \eg clearly highlighted repositories.
We subjectively deduce an overall low availability of source code across \ac{IIDS} research.

Thus, researchers often have to rely solely on the descriptions and evaluation results provided by the paper to verify their code.
Overall, reproducibility is thus challenging as optimally both criteria (public dataset and source code) have to be met.
The increasing use of public datasets promises improvements in at least one direction, while publicly available artifacts accompanying publications remain the exception.

\begin{figure}
	\centering
	\includegraphics[width=\columnwidth]{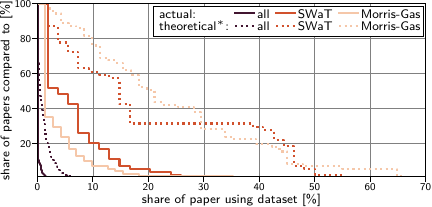}
	$^*$A paper is comparable to all papers that share one dataset and metric and were published at least one year earlier.

	\caption{On average, authors compare \acp{IIDS} to \hlnum{\num{\evalComparesToOnAvg}} approaches from the related work~(black), while theoretically, they could compare to at least \hlnum{\num{\evalCouldCompareToOnAvg}}. This gap increases for papers evaluating the SWaT or Morris-Gas datasets.}

	\label{fig:comparability}
\end{figure}

\subsubsection{Comparability}
\label{sec:eval:comparison:comparability}

Fortunately, cumbersome reproducibility is often not needed when, for example, it suffices to compare results to related work, \eg to prove a novel attack detection approach superior.
This requires that both works have been evaluated on at least one common dataset.
Likewise, to objectively judge their detection performance, both publications must employ at minimum one identical evaluation metric.

To judge the degree of comparability across the research landscape for each publication, we extracted the \emph{actual} number of comparisons made by the authors and calculated the number of \emph{theoretically} possible comparisons.
Therefore, while conducting our \ac{SMS} (\cf \secref{sec:survey}), we gathered how many publications each author uses as comparison references and additionally extracted the exact metrics used in each publication's evaluation.
We estimate the minimum amount of theoretically possible comparisons by counting a publication as comparable if it shares at least one common dataset and metric and was published in an earlier year.
Note that while not every two publications assume the same attack model, comparability can still be justifiable in the cases where the dataset matches since authors should select a dataset that best fits their approach.
This methodology provides a great opportunity to assess actual and theoretical possible comparability, and \figref{fig:comparability} depicts the degree of comparability.

Overall, the number of \emph{actual} comparisons performed by researchers is low, with \hlnum{\num{\evalComparesToOnAvg}} publications on average.
For the two most-common datasets, we observe higher values (SWaT \hlnum{\num{\evalSWaTComparesToOnAvg}} and Morris-Gas \hlnum{\num{\evalMorrisGasComparesToOnAvg}}).
Still, there exists the \emph{theoretical} opportunity for authors to compare a proposed \ac{IIDS} to an average of \hlnum{\num{\evalCouldCompareToOnAvg}} alternatives.
On the one hand, this proves that many works are indeed comparable in terms of datasets and metrics.
On the other hand, prominent datasets help in that regard since their theoretical comparability is higher (SWaT \hlnum{\num{\evalSWaTCouldCompareToOnAvg}} and Morris-Gas \hlnum{\num{\evalMorrisGasCouldCompareToOnAvg}}).
Note that it should not be the ultimate goal to compare against as many publications as possible since quality is preferential before quantity.

Looking closer into the details of \figref{fig:comparability}, it is interesting that \SI{10}{\percent} of the publications evaluating the Morris-Gas dataset (yellow) actually compare only against \hlnum{\SI{7}{\percent}} of different Morris-Gas publications.
However, for SWaT (red), \SI{10}{\percent} of publications are actually compared to about \hlnum{\SI{18}{\percent}} of existing works.
Meanwhile, theoretical comparability for Morris-Gas publications is even higher than for SWaT (dotted lines).
Regarding all publications~(black), a total of \hlnum{\SI{\evalNotComparedToASinglePaper}{\percent}} of publications are not compared to a single \ac{IIDS}.

The state of comparability in the \ac{IIDS} research is decent but with opportunities for improvement in the future as many publications share common datasets and metrics already.

\subsection{Expressiveness of Metrics}
\label{sec:metric:experiment}

Until now, our \ac{SoK} on evaluations of \acp{IIDS} bases on theoretical observations from literature, \eg which datasets and metrics are used.
In the following, we extend our analysis beyond a literature mapping study with \emph{practical} experiments to understand the quantitative impact of metric choices on the evaluation outcomes and to derive metrics that offer high expressiveness.
To this end, we conduct a comparison study across ten \acp{IIDS} from research on two datasets and utilize our evaluation tool (\cf Availability Statement) to compare various metrics.
Especially for newer time series-aware metrics, which are more difficult to compute~\cite{Huetetal2022Local,Hwangetal2022Do}, no common library exists thus far.
Besides the metrics discussed in the following, the tool provides a total of \num{18} point-based and \num{14} time-aware metrics, for which few implementations exist.

\subsubsection{Experiment Design}

As we observed in \secref{sec:eval:datasets}, the \ac{IIDS} research community is governed by two major directions of datasets: network-based datasets such as the Morris-Gas~\cite{Morrisetal2015Industrial} and process data datasets such as SWaT~\cite{Gohetal2016A} containing physical time series data.
We aim to cover both types in our evaluation and thereby also cover two important \ac{IIDS} types from research, namely knowledge- and behavior-based \acp{IIDS} (\cf \secref{sec:bg:iids}).
For knowledge-based \acp{IIDS}, we examine five machine learning approaches~\cite{Perezetal2018Machine,Wijayaetal2020Domain-Based} originally evaluated on the Morris-Gas dataset.
Regarding process data, we leverage five behavior-based \acp{IIDS}, with TABOR basing on timed automata~\cite{Linetal2018TABOR:}, Seq2SeqNN utilizing neural networks~\cite{Kimetal2020Anomaly}, PASAD leveraging singular spectrum analysis~\cite{Aoudietal2018Truth}, SIMPLE implementing minimalistic boundary checks~\cite{Wolsingetal2022Can}, and Invariant mining invariant logical formulas~\cite{Fengetal2019A}.
Contrary to the knowledge-based machine learning approaches, these \acp{IIDS} are evaluated on the temporally \emph{ordered} SWaT dataset, which provides dedicated attack-free training data and testing data, including anomalies.
As an interesting case for the SWaT dataset, we added an \ac{IIDS} that randomly emits alerts by a \SI{50}{\percent} chance.

\subsubsection{Metrics Under Study}

In this study, we focus on the four common point-based metrics accuracy, precision, recall, and F1 (\cf \secref{sec:metrics:utilization}) and modern time series-aware variants of them called enhanced time series-aware recall (eTaPR)~\cite{Hwangetal2022Do} (\cf \appdxref{app:tapr}).
More precisely, eTaP for precision, eTaR for recall, and eTaF1 for F1 (there is no time series-aware accuracy equivalent).
Additionally, we consider the time-aware Affiliation metrics (again expressed as variations of precision, recall, and F1) proposed by Huet \etal~\cite{Huetetal2022Local}, which claim to be robust against randomly generated alerts.
These metrics, like their point-based counterparts, favor high detection rates but diminish the expressiveness of consecutive alarms if they start too early or overhang beyond the duration of an attack.
Furthermore, we examine a variant of $F1$, which allows weighting precision and recall differently.
This may be crucial in industries since cyberattacks are rare compared to normal behavior, preferring a high precision over false alarms.
The datasets in our study already incorporate this class imbalance, with Morris-Gas containing \SI{22}{\percent} malicious data, SWaT just \SI{12}{\percent}, and real deployments likely observing even fewer attacks.
Thus we examine $F_{0.1}$ in addition, weighting precision ten times more than recall.
As the last metric, and since there is only one repetition for each attack type in SWaT, we discuss the percentage of detected scenarios (unique attack types).

\begin{figure}
	\centering
	\subfigure[Morris-Gas]{
		\includegraphics[width=.48\columnwidth, trim=0 0 0 8, clip]{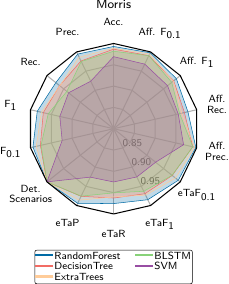}
		\label{fig:practical:left}
	}
	\hspace*{-.2cm}
	\subfigure[SWaT]{
		\includegraphics[width=.48\columnwidth, trim=0 0 0 8, clip]{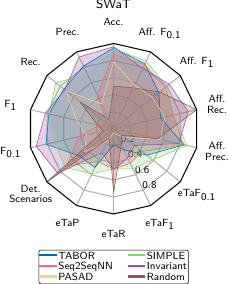}
		\label{fig:practical:right}
	}
	\vspace*{-0.2cm}
	\caption{While point-based metrics rate \acp{IIDS}' performance consistently on the Morris-Gas dataset~\subref{fig:practical:left}, they fail to provide a coherent picture of the time-series dataset SWaT~\subref{fig:practical:right} and judge \acp{IIDS} better than time-aware metrics.}
	\label{fig:practical}
\end{figure}

\begin{figure*}
	\centering
	\includegraphics[width=\textwidth]{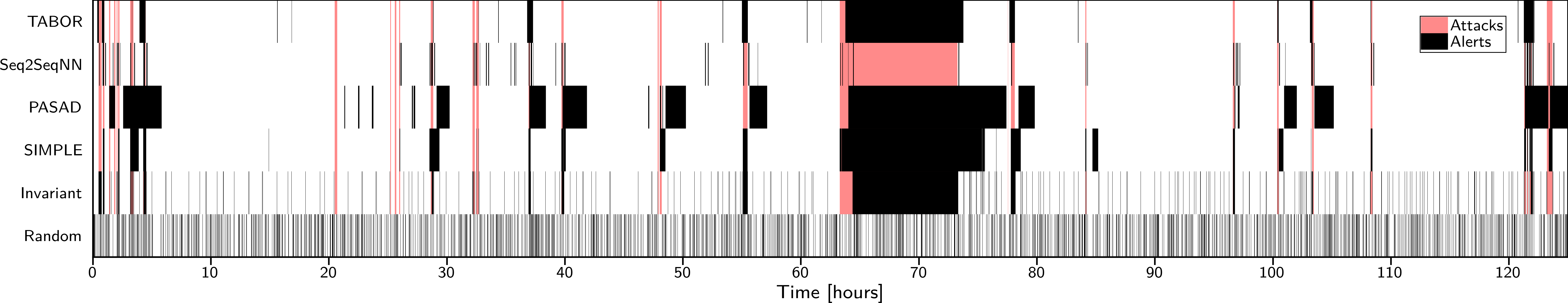}

	\caption{Visualizing alerts side-by-side provides an in-depth view of their distinct alerting behavior. E.g., the underwhelming performance of the Seq2SeqNN \ac{IIDS} in point-based recall (\cf \figref{fig:practical}) can easily be attributed to a single prolonged attack of the SWaT dataset. Note that alerts have been extended to a minimum width of 1 minute for visibility (except for Random).}
	\label{fig:practical:time-series}
\end{figure*}

\subsubsection{Results}

\textit{Point-based.}
We begin with analyzing the knowledge-based \acp{IIDS} on the Morris-Gas dataset in \figref{fig:practical:left}.
Here, the point-based metrics (accuracy, precision, recall, and F1) coherently judge the \acp{IIDS}' performance, \ie one \ac{IIDS} is strictly better than another, and only in recall does the ordering between ExtraTrees and DecisionTrees flip.
The $F_{0.1}$ variant's judgment is in line with the other metrics, likely due to the high amount of malicious samples (\SI{22}{\percent}) in this dataset.
Also, the time series-aware variants draw a nearly identical picture here.
Note that the attack instances of the Morris-Gas dataset correspond to manipulations of individual network packets, and thus temporal effects are minimal.
While the \acp{IIDS} are well at detecting these attacks, it is unclear whether the attacks themselves are actually comprehensible to ones observed on real deployments.
Overall, the considered metrics coherently judge the \acp{IIDS}' performance on the Morris-Gas dataset.

The picture changes for the SWaT dataset comprising of time series of physical states (\cf \tabref{tab:datasets-overview}).
We additionally depict the raw alert emitted by the \acp{IIDS} over time in \figref{fig:practical:time-series}.
First of all, as depicted in \figref{fig:practical:right}, all \acp{IIDS} perform well according to accuracy (more than \hlnum{\num{0.75}}).
Yet, in comparison to all other metrics, accuracy seems to overestimate their capabilities.
We attribute this to SWaT's composition comprising to \SI{12}{\percent} of attacks (which is more realistic than Morris-Gas with \SI{22}{\percent} as attacks are rare in practice), and an \ac{IIDS} that emits no alarms at all would score an accuracy of \hlnum{\num{0.88}} already.

Regarding precision and recall, we observe ambiguity.
Seq2SeqNN falls far behind the other approaches in recall, which we attribute to a single long attack in SWaT (accounting for \SI{63}{\percent} of all attack samples) being missed by the approach (\cf \figref{fig:practical:time-series}).
Besides this attack, Seq2SeqNN achieves decent scores as it correctly detects most of the other attacks (\cf detected scenarios).
Therefore, point-based metrics overvalue this attack, \ie no obvious relation exists between the attack's duration and its severity that would justify this effect.
In contrast to the Morris-Gas dataset, the $F_{0.1}$ score clearly favors \acp{IIDS} with higher precision, and thus TABOR is preferred over the SIMPLE \ac{IIDS} (even though nearly equivalent in F1).

\textit{Time series-aware.}
In general, time series-aware metrics promise to solve these inaccuracies of point-based metrics.
In our practical study, all \acp{IIDS} perform much worse on the time series-aware variants eTa~\cite{Hwangetal2022Do}, which might be the case since they have not been designed for this kind of (potentially more valuable) evaluation.
Here, the SIMPLE \ac{IIDS} is now the best-performing approach according to eTaP, eTaF$_{1}$, and eTaF$_{0.1}$ as its emits alerts are precise, \ie no overshooting as by PASAD or occasional short false-alarms as in TABOR and the Invariant~\ac{IIDS}.
Yet contradicting the traditional recall score, Seq2SeqNN now belongs to the best \acp{IIDS} in the time-series recall pendant (eTaR) probably because the time-aware metric analyzes alarms consecutively, and thus false-negatives of overshooting alarms are not weighted that negatively.

The time-aware affiliation metrics~\cite{Huetetal2022Local} draw a completely different picture since all \acp{IIDS} perform much better.
The ratings for precision, F$_{1}$, and F$_{0.1}$ are mostly consistent, and the \acp{IIDS} only differ significantly in terms of affiliation recall.
However, a random \ac{IIDS}, which the metric should consider as the minimum baseline~\cite{Huetetal2022Local}, is counterintuitively perceived as a better approach than PASAD and TABOR in the affiliation F$_{1}$ score.
In all other point-based and time-aware metrics, this random \ac{IIDS} is perceived as the worst approach (except for detected scenarios and recall).

\textit{False-positive resistance.}
For practical deployment, \acp{IIDS} with many false positives are unsuitable~\cite{Etalle2017From}, and thus identifying those in evaluations is crucial.
In that regard, while the Invariant \ac{IIDS} outperforms all other approaches in many metrics, visually (\cf \figref{fig:practical:time-series}) exhibits the least usable approach due to its plentiful but short-lived false alarms.
Only in the eTa metrics it performs badly.

\subsubsection{Conclusion}

Point-based metrics draw a coherent picture for the Morris-Gas dataset containing a significant amount of attacks with few temporal effects as single network packets were manipulated.
In contrast, authors have to carefully examine their results on the SWaT dataset since, depending on the chosen metric, their \ac{IIDS} may perform excellently or poorly.
These results are in line with Fung \etal~\cite{Fungetal2022Perspectives}, finding that time-series metrics are preferable for reconstruction-based \acp{IIDS} and point-based scores may be misleading.
For the affiliation metrics by Huet \etal~\cite{Huetetal2022Local}, our experiment challenges their results, especially for an \ac{IIDS} that emits alerts randomly.
Thus, a better understanding of how such newer time series-aware metrics have to be interpreted is crucial.

Overall, it is unlikely that a single metric exists that catches all industrial operators' different goals, \eg preferring few false alarms over detected attacks.
\acp{IIDS} should be evaluated with different metrics to truly highlight their capabilities, as cherry-picking metrics may lead to misleading results.
The F$_{0.1}$ score provides an interesting alternative for more realistic scenarios.
Furthermore, visual comparisons exhibit a non-negligible added value to evaluations too.
Lastly, the knowledge- and behavior-based \acp{IIDS} are hardly comparable today since they are divided by dataset type.

\section{Common Issues \& Recommendations}
\label{sec:recommendations}

The huge potential of \acp{IIDS} to combat rising threats from cyberattacks against industrial networks is indisputable.
Unsurprisingly, our systematic analysis (\secref{sec:eval} and \secref{sec:metrics}) shows an unbroken and increasing interest in this research field (\SI{\evalRecentRelativeYearlyIncrease}{\percent} average yearly increase between 2013 and 2021), with at least \num{\evalAcceptedPaper} publications investing great efforts in proposing \acp{IIDS}, which are complemented by further work on creating datasets, designing evaluation metrics as well as surveys and meta-analysis.
However, our \ac{SMS} also unveils and quantifies flaws in this field that hamper scientific progress.
Thus, in the following, we synthesize common issues persisting in \ac{IIDS} evaluations (\secref{sec:recommendations:issues}) and distill recommendations based on the \ac{SMS}'s results (\secref{sec:recommendations:recommendations} and the authors' observations of the research field (\secref{sec:recommendations:discussion}) to move forward to more thorough \ac{IIDS} evaluations and coherent research efforts.

\subsection{Common Issues in IIDS Evaluations}
\label{sec:recommendations:issues}

Our systematic analysis of the \ac{IIDS} research field reveals that the current state-of-the-art \wrt evaluation methodologies has serious inefficiencies, eventually slowing down the overall progress in securing industrial deployments.
Our \ac{SMS}, covering the body of literature until 2021, enables quantifying these inefficiencies and makes (promising) trends visible in contrast to previous meta-surveys and experiments on a usually narrower scale (\cf \secref{sec:bg:rw}).
More precisely, we identify three issues (I1--I3) prevalent in evaluations of \acp{IIDS} and present them along the results from our \ac{SMS} in the following.

\noindent $\blacktriangleright$ \textbf{I1: Dataset Diversity.}
\emph{We identify a lack of diversity in datasets used for evaluations.}
Regarding the utilization of datasets, we find that \acp{IIDS} are evaluated on \num{\evalAverageDatasetsPerPaper} datasets on average (\cf \figref{fig:eval:datasets-per-publication}), which aligns with \num{1.32} datasets on average reported by related work~\cite{Wolsingetal2022IPAL:}.
Notably, the majority (501 publications) considers only a single dataset, despite a significant selection of datasets being publicly available (we identified \num{\evalUniquePublicDatasets} public datasets in our \ac{SMS}, and other work lists \num{23} datasets or \num{61} industrial testbeds~\cite{Contietal2021A}).
Since there exists this large gap between available and utilized datasets, this raises the question of why many datasets are only used rarely.
Possible reasons include datasets being too narrow in scope (\eg focusing on single attack types), too small (providing only few training or testing samples), difficult to use (\eg requiring in-depth knowledge of a specific industrial protocol), or simply not widely known among researchers.
Lua \etal~\cite{Luoetal2021Deep} also find that high-quality datasets are rare.
Moreover, for the few publications that evaluate multiple datasets (\SI{\evalPercentagePaperWithMultipleDatasets}{\percent}), these datasets mostly stem from the same origin (\cf \tabref{tab:dataset-combinations}).
Thus, \acp{IIDS}' evaluations are mostly confined to a single scenario (dataset) and do neither cover the diversity of industrial domains nor communication protocols (\cf \secref{sec:eval:dataset:types}).
Consequently, it remains unclear whether \acp{IIDS} are applicable outside the narrow scenario they have been evaluated in, making real-world deployments risky and requiring repeated efforts for different scenarios.

\noindent $\blacktriangleright$ \textbf{I2: Metrics Ambiguity.}
\emph{Metrics used in evaluations and comparisons pose ambiguity regarding the actual detection performance of \acp{IIDS}.}
Due to the unclear and biased choice of metrics, the \emph{actual} detection performance of proposed \acp{IIDS} often remains unclear, as also claimed by Giraldo \etal~\cite{Giraldoetal2018A}.
Seemingly promising, we observed an increase in the number of utilized metrics (\num{\evalAvgMetricsLatestYear} per publication on average in 2021) while simultaneously moving away from mere textual descriptions (\cf \secref{sec:metrics:utilization}) toward established point-based metrics (\cf \figref{fig:metric-over-time}).
Accuracy, precision, recall, and F1 make up the majority of utilized metrics again~\cite{Luoetal2021Deep}.
However, we also encountered a total of \num{\evalUniqueMetricsBeforeAggregation} flavors of metrics, \eg subtle variations such as multi-class or weighted scores, which further complicates metric ambiguity, but likewise can provide further insights.
Especially in the multi-class classification, \eg when a specific attack should be identified and differentiated from others, researchers leverage diverse constructs often without clearly stating the precise calculations.
This complicates reproduction and comparison since we could not always find out how they were defined, even with significant effort.
At the same time, essential metrics, expected to be provided in combination, are often omitted or incomplete in publications.
I.e., of the \num{\evalPapersWithConfusionMatrix} publications providing the confusion matrix, only \num{\evalAllConfusionAndTopFourMetrics} state accuracy, precision, recall, and F1 in combination (\cf \secref{sec:metrics:utilization:overtime}).
Even more severe, precisely these four point-based metrics, making up \SI{\evalMetricUsageTopFour}{\percent} of the metric usage, do not accurately capture the detection performance in time-series scenarios~\cite{Fungetal2022Perspectives} and are skewed towards the detection of long-lasting attacks (\cf \secref{sec:metric:experiment}).
While plenty new metrics~\cite{Huetetal2022Local,Hwangetal2022Do,Hwangetal2019Time-Series,Tatbuletal2018Precision,Lavinetal2015Evaluating,Gensleretal2014Novel} are designed that supposedly address these issues, these metrics are rarely used in evaluations (only \num{\evalPapersWithTimeaware} publications), likely because a broad understanding about their expressiveness is missing.
Lastly, as our practical experiments show, not a single metric can describe all aspects of an \ac{IIDS}, and visual comparisons can disprove, \eg seeming promising \acp{IIDS}.

\noindent $\blacktriangleright$ \textbf{I3: Underutilized Comparability.}
\emph{Evaluations of \acp{IIDS} do not capitalize on the large potential for comparisons among the vast body of existing research.}
The number of comparisons to related work performed by new \acp{IIDS}' has experienced earlier criticism already~\cite{Giraldoetal2018A}.
On average, an \ac{IIDS} is only compared with \num{\evalComparesToOnAvg} other proposals, slightly more than observed in previous works (0.38)~\cite{Wolsingetal2022IPAL:}.
Yet, in theory, authors could compare an \ac{IIDS} to an average of \num{\evalCouldCompareToOnAvg} other approaches sharing at least one common dataset and metric (\cf \figref{fig:comparability}).
Simultaneously, the current state of the research field leaves researchers large freedom to choose from any of the theoretically suitable publications for their comparisons.
This situation is even aggravated by the sparse commitment to publish artifacts (\cf \secref{sec:eval:comparison:reproducibility}), which leaves researchers no choice other than to reproduce others' works, \eg to ultimately conduct comparability studies---a non-trivial task that is prone to failure~\cite{Erbaetal2020No}.
Meanwhile, researchers have to rely on public datasets and the expressiveness of metrics that both exhibit flaws themselves (\cf I1 and I2).
However, proper comparisons are essential to better understand if and how a novel \ac{IIDS} improves upon existing work and thus collectively move the research field forward.

\subsection{Recommendations for IIDS Evaluations}
\label{sec:recommendations:recommendations}

To address these prevalent issues and thus enhance evaluations as well as the applicability of future \ac{IIDS} research, we extract key aspects from our systematic analysis and turn them into six actionable and practical recommendations (R1--R6).

Since our recommendations target different parties involved in \ac{IIDS} research, we address them to
(i) \emph{researchers} designing new detection approaches, evaluating them, and comparing them to the state-of-the-art;
(ii) dataset \emph{creators} recording qualitative datasets or providing simulations and testbeds; and
(iii) industrial \emph{operators} with precise knowledge of the individual needs of \acp{ICS}' striving to role out \acp{IIDS} in practice.

\noindent $\blacktriangleright$ \textbf{R1: Evaluate More and Diverse Datasets.}
\emph{Researchers} should use the many readily available datasets to comprehensively evaluate their \acp{IIDS} for different industrial domains, communication protocols, and attack types.
Using multiple, especially diverse datasets avoids overfitting~\cite{Wolsingetal2022Can}, boosts generalizability across \ac{ICS}, enables insights across multiple domains, and allows assessing the potential efforts required to facilitate (widespread) deployability across industries.
For a concise dataset selection, we recommend focusing on publicly available datasets such as those listed in Conti~\etal's~\cite{Contietal2021A} comprehensive datasets and testbeds overview.
For evaluations requiring process data, datasets of multiple origins and industrial domains should be used.
Likewise, for \acp{IIDS} operating on network traffic, generalizing the approach to different industrial protocols should be considered.
Moreover, specialized datasets that, \eg model a single attack type, cover a niche industrial domain, or deploy rarely used protocol, still provide substantial added value when used in combination with other, more general datasets to better understand the capabilities and limitations of an \ac{IIDS}.
Additionally, researchers can consider datasets containing attacks and faults (\eg the IEC61850SecurityDataset~\cite{Biswasetal2019A}) to evaluate whether their proposed \acp{IIDS} can differentiate these kinds of unwanted behavior to facilitate swift and correct reactions by operators to alerts.
Lastly, to ease evaluations on a multitude of datasets with potentially varying formats, agreeing on unified dataformats, such as IPAL~\cite{Wolsingetal2022IPAL:}, may help lower the burdens for researchers.

\noindent $\blacktriangleright$ \textbf{R2: Provide High Quality Datasets.}
Dataset \emph{creators} should provide the research community with high-quality and diverse datasets to counteract the current bias to two major datasets (\cf \tabref{tab:datasets-overview}).
To ensure the practical relevance of datasets, they should ideally be generated in close cooperation with industrial partners~\cite{Pennekampetal2021Collaboration} since otherwise, \acp{IIDS} designed upon them risk not being of practical use to industrial \emph{operators}.
Such collaborations, even though costly~\cite{Apruzzeseetal2022SoK:}, also allow enriching datasets with properties and demands of actual industrial deployments, \eg the criticality of an attack, an acceptable delay until which a detection is excepted, or documentation of how long the \ac{ICS} behaves abnormally after an attack until it stabilized again.
Furthermore, research lacks datasets that tackle the needs of all \ac{IIDS} flavors (\cf \secref{sec:eval:dataset:types}) simultaneously.
For one, only a few datasets (Faramondi~\etal providing a rare exception~\cite{Faramondietal2021A}) already combine network traffic and process data, which is necessary to compare \acp{IIDS} that work on these different data types.
Providing captures for all data types in a single dataset, or even possibly from multiple vantage-points in the (network) topology, likewise enables assessing the effectiveness of (combining) network- and process-based approaches, whilst consolidating the overall research landscape.
Moreover, datasets should be designed and created such that they are applicable to both knowledge-based and behavior-based \ac{IIDS} training (currently, no corresponding dataset is known to us), \eg by including repetitions and variations of the same attack, providing sufficient long samples of benign behavior, and including novel attacks, which are not previously trained on, to avoid the drawing of false conclusions~\cite{Kusetal2022A}.
Another crucial factor is a high quality labelling of the dataset, which can be difficult do perform right~\cite{Wolsingetal2022Can}, but is of utmost importance to accurately calculate a given metric.
Lastly, reviews of datasets' quality~\cite{Lanferetal2023Leveraging}, \eg with statistical means accessing the data distribution and stability~\cite{Turrinetal2020A} or analysis of how easily a dataset can be ``solved''~\cite{Wolsingetal2022Can}, can guide developers in choosing a relevant dataset.
For more concrete advice on how scientific \ac{IIDS} evaluation datasets should be designed, please refer to the works by G{\'o}mez \etal~\cite{Perales-Gomezetal2019On} and Mitseva \etal~\cite{Mitsevaetal2023Challenges}.

\noindent $\blacktriangleright$ \textbf{R3: Use Standardized and Accessible Metrics.}
\emph{Researchers} should carefully consider the use of metrics and rely on both common (flawed) metrics for comparability as well as recent time-series aware metrics~(\cf \secref{sec:metrics:utilization}) that attempt to mitigate known flaws.
In that regard, meta-studies on how metrics fare against each other, as done in \secref{sec:metric:experiment} and by Huet \etal~\cite{Huetetal2022Local}, or in-depth analysis of individual metrics, such as the F1 score~\cite{Christenetal2024A}, help understand the expressiveness of evaluations.
Ideally, a wide variety of different metrics is used to disseminate the performance of newly proposed \ac{IIDS}, as also recommended by Umer \etal~\cite{Umeretal2022Machine}, which would also facilitate comparisons in the future.
Especially with the rise of new metrics and to standardize the evaluation process, \emph{researchers} should be equipped with adequate tooling to calculate these metrics easily.
Our evaluation tool used in \secref{sec:metric:experiment} and published along this paper will greatly help in that regard.
To facilitate a sensible choice of metrics and ensure comparability of related \ac{IIDS} approaches, dataset \emph{creators} should explicitly define standard evaluation metrics for their datasets, as has been done, \eg for the HAI dataset~\cite{Shinetal2020HAI}.
First, fixing metrics a priori ensures the neutrality of evaluations and reduces potential biases in their selection by researchers.
More importantly, however, dataset developers know the underlying \ac{ICS} best, \eg \wrt the impact of false positives or the likelihood of attacks.
Often they are the only people with the necessary expertise to identify the demands of a cybersecurity solution and, thus, the most valuable metrics to benchmark an \acp{IIDS} in their scenario.

\noindent $\blacktriangleright$ \textbf{R4: Facilitate Comparability With Public Artifacts.}
\emph{Researchers} should make the artifacts publicly available~\cite{Etalle2017From}, especially \ac{IIDS} implementations, underlying their work to facilitate comparability of \ac{IIDS} research.
If artifacts cannot be provided, \eg due to licensing issues or private datasets, we recommend that researchers at least release the precise \ac{IIDS} outputs, \eg a list with all packets classified as malicious by an \ac{IIDS}.
These outputs, together with the (anonymized) labels of the dataset, suffice to calculate further metrics retrospectively, thus gaining new insights into the \ac{IIDS}'s performance even after publication.
It also facilitates reconstructing the metrics' exact definitions if not stated in the paper.
Publishing the alerts when evaluated on a public dataset is valuable if published alongside its implementation, as getting research code to run and produce the same results independently is often hard work (\eg due to lacking documentation), especially some years down the road.
Such published labels directly avoid the current lock-in to metrics during the time of publication, thus greatly enhancing the comparability of \ac{IIDS} research.
This freedom is especially crucial in an early stages of \ac{IIDS} research since it is unknown which metrics and evaluation methodologies will eventually gain acceptance.

While these recommendations endorse coherent research and evaluations, we want to highlight that innovative advancements in (\ac{IIDS}) research may need to deviate from the ordinary and excel established research standards and procedures.

\subsection{Further Recommendations and Discussion}
\label{sec:recommendations:discussion}

Besides the previous four recommendations grounded in our \ac{SMS}'s results, we identified two further aspects that we demand as crucial for the success of \acp{IIDS} toward deployments in actual \ac{ICS} to strengthen their cybersecurity.

\noindent $\blacktriangleright$ \textbf{R5: Strive for Continuous Feedback Loops.}
\emph{All} stakeholders should strive for coherence and applicability of \ac{IIDS} research.
\emph{Researchers} should avoid proposing isolated \acp{IIDS} without proving their necessity and bridge the gaps between related branches for greater coherence~\cite{Wolsingetal2022IPAL:}.
At the same time, meta-surveys that critically review the state-of-the-art have to provide directions regarding which approaches work well in given settings, which datasets and metrics are suitable, and which approaches should \acp{IIDS} should compare to.
In that regard, new evaluation methodologies, such as pragmatic assessments for machine learning-based network intrusion detection systems, are already being developed~\cite{Apruzzeseetal2023SoK:}.
Lastly, a continuous exchange between all stakeholders should be established~\cite{Pennekampetal2021Collaboration}, \eg in the form of public talks, workshops, or the dissemination of scientific publications.
Only then can industrial \emph{operators} stay informed about recent advancements and likewise keep dataset \emph{creators} updated to ensure overall research strives for practical applicability.
As an initial step in that direction, we provide the artifacts of our broad \ac{SMS}, which can serve as the foundation for future surveys on more specific topics, such as in-depth analyses of the proposed detection methodologies or benefits and drawbacks of the wide variety of (newly proposed) evaluation metrics.

\noindent $\blacktriangleright$ \textbf{R6: Think Beyond Alerting.}
As already shortly outlined in \secref{sec:bg:iids} and as depicted in \figref{fig:iids:dimensions}, \emph{researchers} should extend their focus beyond optimal attack detection coverage and on the required actions after \ac{IIDS} alerts.
Such actions may include steps to reduce the number of false alerts by fusing multiple \acp{IIDS}~\cite{Wolsingetal2023One}, enhance the understandability of alert~\cite{Etalle2017From,Sengetal2022Why}, localize the attacker~\cite{Al-abassietal2022A}, mitigate an attack's damage potential~\cite{Sunetal2021SoK:}, recover the system to a safe state~\cite{Weietal2020Cyber-Attack}, and lastly, perform forensics to learn for the future~\cite{Kebande2022Industrial}.
Given this chain of tasks operators have to execute, which may include temporal interruptions of the process, it may also be crucial for researchers to consider the costs of (false) alarms emitted by their solutions.
While research on follow-up procedure of \ac{IIDS} alerts is currently critically underrepresented in the literature, this is partially caused by the secrecy of industrial \textit{operators}.
The sharing of detailed information about the operation of real-world \acp{ICS} allows researchers to propose valuable and actionable improvements to current processes.
Moreover, this information also allows researchers to design suitable evaluation methodologies to evaluate the performance of the processes following an alarm.
Overall, \ac{IIDS} should thus no longer be considered as an isolated system, but the step from detection to (incident) response should be considered a tightly interlocked process.

In conclusion, throughout all recommendations R1--R6, our \ac{SoK} advocates for more coherent and practical \ac{IIDS} research comprising all parties involved in establishing \ac{ICS} security.
Yet, precise actions to accomplish these goals are not directly apparent from such a broad \ac{SMS} and may demand further investigation into more niece topics refining our results.

\section{Conclusion}
\label{sec:conclusion}

The ongoing digitization of industries and increasing exposure of \ac{ICS} to the Internet are accompanied by a rise in cyberattacks.
Consequently, the new research field of industrial intrusion detection, promising to provide an easily deployable solution to uncover even sophisticated attacks, gained traction.
In 2021 alone, \num{\evalPubsLatestYear} new detection approaches were proposed.

This \ac{SoK} presents the first systematic attempt to shed light on this fast-growing research field and how different approaches are evaluated.
Our thorough analysis of \num{\evalAcceptedPaper} publications reveals the tremendous efforts invested by the community to protect industrial systems.
However, when it comes to evaluating detection approaches, we uncover widespread issues \wrt dataset diversity, the ambiguity of metrics, and missed opportunities for comparability, hampering the overall progress of this quickly growing research field.
Based on our systematic analysis, we formulate actionable recommendations to overcome these issues and thus
bring the entire research domain forward to sustainably and significantly improve the security of (real-world) industrial deployments.

\begin{acks}
Funded by the Deutsche Forschungsgemeinschaft (DFG, German Research Foundation) under Germany's Excellence Strategy -- EXC-2023 Internet of Production -- 390621612 and by the German Federal Office for Information Security (BSI) under project funding reference number 01MO23016D (5G-Sierra).
The responsibility for the content of this publication lies with the authors.
The authors would like to thank Antoine Saillard, and Frederik Basels for their work on contributing to the evaluation tool.
\end{acks}

\bibliographystyle{plainurl}

\appendix

\section{Systematic Literature Review}
\label{sec:appx:slr}

To find relevant literature proposing \acp{IIDS} with the help of various search engines, we derived a search string during the design of the \ac{SMS} (\cf \secref{sec:survey}).
As depicted in \figref{fig:survey-design}, the search string combines collections of keywords for the phrases \emph{industrial} and \emph{detection}.
After validating the outcome of several search strings and keywords against known literature of that research landscape, we derived the following search string utilized in the \ac{SMS}:

\begin{center}
\textit{(``Industrial Control Systems'' OR ``Process Control Systems'' OR ``Supervisory Control and Data Acquisition'' OR \\ ``PCS'' OR ``ICS'' OR ``SCADA'') \\
AND \\
(``Attack Detection'' OR "Anomaly Detection'' OR ``Intrusion Detection'')}
\end{center}

The search string was applied to the combination of titles, abstracts, and keyword of all publications.
Note that search engines do not care about capitalization or singular vs plural.

\section{Overview on Utilized Metrics}
\label{sec:appx:metrics}

Detection performance metrics quantify the capabilities of an \ac{IIDS}, \ie its ability to differentiate benign from malicious behavior.
Thus, such metrics are essential to achieve objective comparisons among publications in research.
Aside from subjective textual descriptions, we found a large variety of different metrics and flavors throughout our \ac{SMS}, ranging from well-defined and widely adopted ones to novel proposals.
To facilitate a systematic overview, we have derived a taxonomy in~\secref{sec:metric:taxonomy}.
In the following, we now provide additional details about the most important metrics,
covering well-established point-based metrics~(\secref{sec:appx:metrics:pbm}) and promising time series-aware metrics~(\secref{app:time-series-aware-metrics}).
Please refer to \tabref{tab:metric:taxonomy} for synonyms by which these metrics are also known.

\subsection{Point-based Metrics}
\label{sec:appx:metrics:pbm}

The ``traditional'' way to evaluate \acp{IIDS} is to utilize a benchmarking dataset which includes a label for each entry in the dataset, stating whether this entry is benign or malicious, \ie it corresponds to a~(specific) cyberattack.
Note that multi-class \acp{IIDS} and corresponding metrics also exist, \eg \acp{IIDS} that precisely identify and attribute each conducted cyberattack type.
However, for the sake of simplicity, we only refer to binary (malicious and benign) metrics in the following.

Given the labels of a dataset and the outputs/alerts of an \ac{IIDS}, one can compare them point-to-point to estimate in how many instances the output is correct~(\emph{T}) and how often it is false~(\emph{F}), \ie deviating from the expected label.
Various metrics then derive performance scores with different meanings, usually normalization in the interval of $[0, 1]$.
For a detailed discussion on each metric, please refer to~\cite{Powers2011Evaluation:, Arpetal2022Dos}.

\subsubsection{Confusion Matrix}
\label{app:confusion}

Beginning with the dataset's labels and the \ac{IIDS}'s alarms, four different outcomes are possible:
First, true negatives~(\emph{TN}) are all instances where the dataset label is benign, and the \ac{IIDS} has not raised an alarm.
Likewise, true positives~(\emph{TP}) correspond to those attack instances within the dataset that are correctly identified as attacks.
In contrast, false negatives~(\emph{FN}) are attack instances that are incorrectly not detected by the \ac{IIDS}.
Lastly, false positives~(\emph{FP}) are false alarms triggered even though no attack has occurred.

Counting all of these four possible outcomes across a dataset yields the confusion matrix, laying the foundation for many point-based metrics introduced in the following.

\subsubsection{Recall / True Positive Rate (TPR)}
\label{app:tpr}

This metric states how many attacks of the dataset are actually detected by an \ac{IIDS}.
Naturally, an \ac{IIDS} has to detect as many attack instances of a dataset as possible.

$$\frac{TP}{TP + FN}$$

\subsubsection{Miss-Rate / False Negative Rate (FNR)}
\label{app:fnr}

In contrast to TPR, FNR measures the fraction of missed attacks.
Hence, a lower score is preferred.

$$\frac{FN}{TP + FN}$$

\subsubsection{Specificity / True Negative Rate (TNR)}
\label{app:tnr}

Since cyberattacks are rare, it is crucial that an \ac{IIDS} does not trigger alarms during benign system behavior.
Thus, TNR defines the fraction of correctly classified benign behavior.
A high TNR score is preferential.

$$\frac{TN}{TN + FP}$$

\subsubsection{Fall-out / False Positive Rate (FPR)}
\label{app:fpr}

Similar to TNR, \acp{IIDS} should only trigger an alarm in case of actual attacks.
Therefore, FPR calculates the fraction of false alarms across the dataset, which has to be as low as possible.

$$\frac{FP}{FP + TN}$$

\subsubsection{Precision / Positive Prediction Value (PPV)}
\label{app:ppv}

When focusing on the alarms triggered by an \ac{IIDS}, PPV defines the fraction of correctly detected attacks among all existing attacks in the dataset.

$$\frac{TP}{TP + FP}$$

\subsubsection{Negative Prediction Value (NPV)}
\label{app:npv}

Contradicting the PPV metric, NPV counts the number of correctly classified negative predictions among all attack free parts in the dataset.

$$\frac{TN}{TN + FN}$$

\subsubsection{Accuracy}
\label{app:acc}

The first metric capturing the overall number of correct classifications is accuracy.
The higher the accuracy score is, the more reliable the predictions of the \ac{IIDS} are.

$$\frac{TP + TN}{TP + TN + FP + FN}$$

\subsubsection{F1}
\label{app:f1}

For intrusion detection, there is an inherent tradeoff between achieving a maximal number of detected attacks~(TPR) while reducing false positives~(expressed by PPV as correct alarms).
The F1~score combines both design goals into a single metric through the harmonic mean.
Note that F~scores with different TPR and PPV weightings exist, as discussed in \secref{sec:metric:experiment}.

$$\frac{2 TP}{2 TP + FP + FN}$$

\subsubsection{Receiver operating Characteristics Curve (RoC)}
\label{app:roc}

\acp{IIDS} and their detection models may require fine-tuning hyperparameters, \eg to determine a threshold upon which an alert is triggered.
Since the previous metrics evaluate \acp{IIDS} for a fixed setting, it is impossible to describe their behavior across the parameter range.

The RoC curve is a method to visualize multiple \ac{IIDS} configurations and their performance by plotting FPR on the x-axis and TPR on the y-axis.
Since each entry represents a tradeoff for a specific \ac{IIDS} model, the RoC curve enables developers to choose a suitable configuration visually.
For a detailed discussion on the appropriate usage of RoC curves, please refer to Arp \etal~\cite{Arpetal2022Dos}.

\subsubsection{Area under Curve (AuC)}
\label{app:auc}

The Area under~(the RoC) Curve abstracts from a visual performance indicator and defines a quantitative metric expressing \ac{IIDS} performance for a variety of configurations by integrating the enclosed area.
If only a single configuration is measured with the RoC visualization, AuC can be simplified into to the following formula~\cite{Powers2011Evaluation:}.

$$1- \frac{\textit{FPR + FNR}}{2}$$

\subsection{Time Series-aware Metrics}
\label{app:time-series-aware-metrics}

Besides point-to-point comparisons between dataset labels and \ac{IIDS} alarms, recent metrics strive towards time series-awareness~\cite{Kimetal2022A,Hwangetal2019Time-Series,Hwangetal2022Do,Gargetal2022An,Jamesetal2020Novel,Lavinetal2015Evaluating,Tatbuletal2018Precision}.
They usually define attacks and alarms as a continuous time range with start and end points.
Alarm intervals should be largely overlapping, \ie an alarm is expected immediately after the start of an attack and should stop in time after the attack phase.

\subsubsection{Detected Scenarios}
\label{app:detscen}

In contrast to point-based metrics and since time series-aware attacks are considered a single instance, it suffices to be indicated by an \ac{IIDS} with a single short alarm.
Unlike point-based metrics, attacks are considered a single instance in the time series-aware domain, and thus it is sufficient for an \ac{IIDS} to trigger a single alarm.
Therefore, detected scenarios enumerates the number of independent attack instances detected by at least a single alarm.

\subsubsection{Detection Delay}
\label{app:detdelay}

Nonetheless, early detection is still preferential in time-critical scenarios as this increases the time to respond to an attack.
Thus, for all attacks in the dataset, the detection delay aggregates the time intervals between the start of an attack and the time of the first detection.

\subsubsection{Enhanced Time-aware Precision and Recall}
\label{app:tapr}

Recently in 2022, Hwang \etal~\cite{Hwangetal2022Do} proposed their (enhanced) time series-aware variants for classical point-based metrics, \ie precision, recall, and F1, addressing known issues when adopting point-based metrics to time series-aware evaluations.
For instance, while point-based recall weights long attacks as more important, the new time series-aware recall variant (eTaR) treats all consecutive attacks equally.
To replace precision, eTaP implements diminishing returns for long-lasting alarms.
Lastly, the new proposed eTaF~score is defined in the same way as the regular F~score~(\cf \secref{app:f1}) but leverages the substitute eTaP and eTaR metrics.

\subsubsection{Affiliation Metrics}
\label{app:affiliation}

A similar approach was taken by Huet \etal~\cite{Huetetal2022Local} with their affiliation metrics.
Again they consider alerts and the ground truth as continuous time-ranges instead of independent points.
In the first step, their approach associates each alert to the closest ground truth, called local affiliation, and then calculates the individual distances for precision and recall.
What makes their approach interesting is, that the final result is normalized in comparison to an \ac{IIDS} that emits alerts at random.
As also done for eTaF1, the time-aware affiliation precision and recall variants are averaged into the affiliation F score.

\end{document}